\documentclass[sigconf]{acmart}
\AtBeginDocument{%
  }

\usepackage{subcaption}

\setcopyright{acmlicensed}
\copyrightyear{2026}
\acmYear{2026}
\acmDOI{XXXXXXX.XXXXXXX}
\acmConference[Conference acronym 'XX]{}{}{}  
\acmISBN{XXX-X-XXXX-XXXX-X/2026/05}

\begin{document}

\title{Revisiting JBShield: Breaking and Rebuilding Representation-Level Jailbreak Defenses}

\author{Kemal Derya}
\affiliation{%
  \institution{Worcester Polytechnic Institute}
  \city{Worcester}
  \state{MA}
  \country{USA}
}
\email{kderya@wpi.edu}

\author{Berk Sunar}
\affiliation{%
  \institution{Worcester Polytechnic Institute}
  \city{Worcester}
  \state{MA}
  \country{USA}
}
\email{sunar@wpi.edu}








\renewcommand{\shortauthors}{Derya et al.}

\begin{abstract}

Defending large language models (LLMs) against jailbreak attacks, such as Greedy Coordinate Gradient (GCG), remains a challenge, particularly under adaptive threat models where an attacker directly targets the defense mechanism. JBShield, a recent jailbreak defense with a 0\% attack success rate (ASR) in some settings, detects malicious prompts via two concept signals, a \textit{toxic} concept and a \textit{jailbreak} concept. It flags a prompt only when both concept scores exceed their respective thresholds, forming an AND-gate. We show that its robustness does not hold under adaptive optimization. We design JB-GCG, which modifies GCG's objective to combine two terms: refusal-direction suppression via cosine similarity between the refusal direction and hidden-state representations, and toxic-concept regularization via JBShield's own toxic concept score. JB-GCG consistently drives the toxic concept below threshold while leaving the jailbreak concept active, effectively breaking JBShield's detection structure. Across five configurations on Llama-3-8B, JB-GCG achieves an average ASR of 46.2\%, reaching up to 53.4\% in the strongest setting. We further evaluate the full JBShield pipeline and show that our attack remains effective against JBShield-M, achieving ASR up to 30.7\% across evaluated settings. The attack persists across multiple JBShield recalibrations ($N{=}10$--$50$), confirming that the vulnerability is structural rather than calibration-specific.

To understand why adaptive GCG-style attacks succeed, we analyze the cosine-similarity signatures of jailbreak representations and find that they occupy a distinctive region in refusal-direction fingerprint space that neither harmless nor harmful prompts inhabit. Motivated by this observation, we introduce \textbf{Representation Trajectory Verification} (RTV), a new defense based on Mahalanobis outlier detection over multi-layer refusal-direction fingerprints, requiring only harmless and harmful prompts for calibration. RTV attains an AUROC of 0.99 against our JB-GCG attack. Finally, we design and evaluate an additional adaptive attack against RTV with full white-box knowledge of the defense; even under this stronger threat model, the best attack achieves only 7\% ASR at much increased $13\times$ the computational cost. Our results show that strong non-adaptive detection does not imply robustness under adaptive threat models, and that multi-layer representation consistency is a more reliable foundation for jailbreak detection than single-layer concept similarity. The code will be available after the reviews.

\end{abstract}




\maketitle

\section{Introduction}

Large language models (LLMs) are used for text generation~\cite{brown2020language, openai2023gpt4, touvron2023llama}, instruction following~\cite{ouyang2022training, wei2022finetuned, sanh2022multitask}, and external tool use~\cite{schick2023toolformer, yao2023react, patil2024gorilla}. To reduce misuse, these models are typically aligned using safety fine-tuning methods such as reinforcement learning from human feedback (RLHF)~\cite{christiano2017deep, stiennon2020learning, ouyang2022training}. Despite these safeguards, jailbreak attacks remain effective and carefully crafted prompts can cause aligned models to ignore safety policies and produce unsafe outputs~\cite{zou2023universal, wei2023jailbroken, chao2025jailbreaking, liu2024autodan, perez2022red, shen2024donowllm}. To strengthen LLM safety beyond internal alignment, a growing body of external defenses spans guard models~\cite{inan2023llamaguard, zeng2024shieldgemma, han2024wildguard}, input-side filtering~\cite{robey2023smoothllm, alon2023detecting, jain2023baseline, kumar2024certifying, xie2023defending}, decoding-time intervention~\cite{xu2024safedecoding, li2024rain, phute2023llm}, and representation-level analysis and representation rerouting~\cite{zhang2025jbshield, xie2024gradsafe, zou2024improving}, yet each commits to a single decision point that adaptive adversaries can target.

Despite the spread of external defenses, they exhibit structural limitations that motivate a different approach. Guard models such as Llama Guard~\cite{inan2023llamaguard}, ShieldGemma~\cite{zeng2024shieldgemma}, and WildGuard~\cite{han2024wildguard} treat detection as a surface-level text classification problem, leaving them blind to adversarial suffixes that are semantically meaningless yet behaviorally effective~\cite{adiletta2025super, zou2023universal}. Input-side defenses such as SmoothLLM~\cite{robey2023smoothllm} and perplexity filtering~\cite{alon2023detecting,jain2023baseline} rely on unnatural text properties of known attacks that adaptive attackers can circumvent by generating semantically coherent jailbreaks~\cite{liu2024autodan,chao2025jailbreaking}. Decoding-time interventions such as SafeDecoding~\cite{xu2024safedecoding} and RAIN~\cite{li2024rain} depend on the safety signals that jailbreaks are designed to suppress. Representation-level defenses come closest to addressing the root cause, but collapse the model's layer-wise computation into a single decision point. JBShield~\cite{zhang2025jbshield} inspects two concept subspaces at a single frozen layer via an AND-gate and GradSafe~\cite{xie2024gradsafe} reduces the full forward pass to a scalar gradient similarity.

Arditi~\emph{et~al.}~\cite{arditi2024refusal} show that refusal in aligned LLMs is mediated by a single direction in the residual stream, and that adversarial suffixes such as GCG operate by suppressing propagation along this direction. This makes refusal-direction alignment a natural detection feature, as demonstrated by HiddenDetect~\cite{jiang2025hiddendetect}, which scores inputs by cosine similarity with a refusal vector at selected layers. Cosine-similarity detectors such as HiddenDetect aggregate to a scalar score rather than treating the multi-layer profile as a distributional object. Separately, Mahalanobis distance is a well-established primitive for out-of-distribution (OOD) and has proven useful in adversarial example detection in deep feature spaces~\cite{lee2018simple, podolskiy2021revisiting}. It has recently been applied to LLM safety tasks as well~\cite{hua2025rethinking, feng2026benchmarking, li2026bolster}. Prior Mahalanobis-based safety detectors~\cite{hua2025rethinking, feng2026benchmarking, li2026bolster} operate on raw hidden states at a single layer and token position, and no prior work combines refusal-direction cosine similarities across layers and positions with Mahalanobis outlier scoring.

Any representation-level defense must be evaluated against adaptive attacks. Bailey~\emph{et~al.}~\cite{bailey2026obfuscated} demonstrate that state-of-the-art representation-space defenses, including Mahalanobis-based OOD detection on LLM activations, can be bypassed by obfuscated activation attacks. Schwinn and Geisler~\cite{schwinn2024revisiting} report a similar result against Circuit Breakers~\cite{zou2024improving} with only three modifications to the embedding-space attack used in the original evaluation. These results indicate that sheer dimensionality of the detection surface is not the relevant source of robustness, since Mahalanobis scoring over raw LLM activations operates on thousands of dimensions.

Among representation-level defenses, we focus on JBShield which uses cosine similarity between hidden-state representations and calibrated \textit{toxic} and \textit{jailbreak} concept directions. It reports $0\%$ ASR against GCG-based attacks on Llama-3-8B in which the attack objective is modified to weaken the toxic concept and enhance the jailbreak concept. In a realistic adaptive threat model, the attacker has knowledge of the deployed defense and can modify GCG's objective to optimize for detector evasion. Motivated by this, we revisit JBShield under an adaptive threat model where we modify GCG's objective via three successive methods, each informed by the limitations of the previous one. We assess each method's attack success and ultimately arrive at a tailored GCG-style adversarial suffix attack (JB-GCG) that both induces unsafe model behavior and evades JBShield's detector. While straightforward GCG-style variants are insufficient to bypass the defense consistently, this analysis guides the design of a stronger adaptive attack. The resulting adversarial prompts expose a near-zero distinctive zone in refusal-direction fingerprint space that JBShield's single-layer inspection entirely misses.

Building on this observation, we propose \emph{Representation Trajectory Verification} (RTV), which computes cosine similarity to the refusal direction across multiple layers and multiple token positions to construct a fingerprint, and applies Mahalanobis distance as the outlier score on this fingerprint. RTV ties a mechanistically grounded attack signature, i.e. GCG's suppression of the refusal direction~\cite{arditi2024refusal}, to a detection framework with well-characterized statistical behavior. Moreover, RTV avoids both the single-layer, single-position bottleneck of prior Mahalanobis-based safety detectors and the single-feature, single-position limitation of cosine-similarity detectors. Given the adaptive-robustness concerns raised above, we do not claim that RTV's fingerprint is intrinsically harder to obfuscate by virtue of its size. Instead, we hypothesize that its robustness, if any, derives from two structural properties. First, an adaptive attack must simultaneously \emph{suppress} refusal-direction propagation (to jailbreak the model) and \emph{preserve} the fingerprint of harmless inputs (to evade detection), which is a tension absent from raw-activation detectors. Second, the fingerprint imposes consistency constraints across multiple layers and token positions rather than at a single layer. Whether an adaptive attacker can resolve these tensions with a modified loss at comparable cost is an empirical question, and we address it through adaptive evaluation in Section~\ref{sec:rtv_adaptive}.

\subsection{Contributions}
\begin{enumerate}
    \item We provide an analysis of JBShield's detection mechanism and show that its AND-gate and single-layer concept extraction pose a structural vulnerability exploitable by an adaptive attack.
    \item We introduce JB-GCG, an adaptive attack that modifies GCG's objective to combine refusal-direction suppression with toxic-concept regularization, achieving up to $53.4\%$ ASR against JBShield-D and $30.7\%$ against JBShield-M on Llama-3-8B, in settings where JBShield reports $0\%$. JB-GCG persists across different JBShield configurations ($N{=}10$--$50$), which confirms the structural vulnerability.
    \item We characterize the multi-layer representation-level signature of JB-GCG, showing that adversarial prompts produce layer-inconsistent refusal-direction profiles that are absent from harmless and harmful prompts.
    \item We propose RTV, a defense based on Mahalanobis outlier detection over a 15-dimensional refusal-direction fingerprint, achieving $0.99$ AUROC against JB-GCG without requiring jailbreak examples for calibration.
    \item We evaluate RTV under adaptive attack with full white-box knowledge, showing that the strongest attacker achieves only $7\%$ ASR at $13\times$ computational cost, where evasion is bounded by the conflict between refusal suppression and fingerprint evasion.
\end{enumerate}

\section{Related Work}
\label{sec:related_work}

\subsection{External Jailbreak Defenses}
\label{sec:related_defenses}

External defenses for aligned LLMs differ in where they intervene and what signal they consume. Llama Guard~\cite{inan2023llamaguard} fine-tunes a model on a custom safety taxonomy. ShieldGemma~\cite{zeng2024shieldgemma} scales this to a family of task-specific detectors with output-side coverage. WildGuard~\cite{han2024wildguard} extends the taxonomy to adversarial and benign-refusal edge cases. All three share a common limitation where their training signal is surface-level text, which leaves them blind to semantically meaningless suffixes such as those produced by GCG~\cite{zou2023universal, adiletta2025super}.

Input-side defenses intervene before the prompt reaches the model. Perplexity filtering~\cite{alon2023detecting, jain2023baseline} rejects inputs whose language-model perplexity under a reference model exceeds a threshold, targeting the unnatural token distributions of early adversarial suffixes. Erase-and-check~\cite{kumar2024certifying} provides certified robustness by verifying that the model refuses every token-level substring of the input. These defenses are effective against attacks whose signatures are preserved in the text itself, but adaptive attackers can circumvent them by optimizing for semantically fluent jailbreaks~\cite{liu2024autodan, chao2025jailbreaking}.

Decoding-time defenses shift the intervention to generation. SafeDecoding~\cite{xu2024safedecoding} trains an auxiliary expert model on (harmful, refusal) pairs and uses it to re-weight the token distribution of the target model during sampling. RAIN~\cite{li2024rain} introduces a self-evaluation loop that rewinds generation when the model's own critique flags its output as unsafe. Self-defense~\cite{phute2023llm} asks a second LLM to audit completions post hoc. These approaches depend on the target model's own safety signals remaining intact under attack, the very signals GCG-style attacks are designed to suppress~\cite{arditi2024refusal}.

Representation-level defenses operate on hidden-state signals. GradSafe~\cite{xie2024gradsafe} computes the cosine similarity between the target model's gradient with respect to a harmful-response target and a reference gradient computed on known harmful prompts, producing a scalar score. Gradient Cuff~\cite{hu2024gradient} formalizes a refusal-loss function and exploits the observation that its gradient norm is larger on malicious queries than on benign ones, using a two-stage test over function value and gradient magnitude. JBShield~\cite{zhang2025jbshield} inspects two rank-1 concept subspaces (toxic and jailbreak) at a single frozen layer via an AND-gate, detecting prompts that activate both concepts simultaneously. Alert~\cite{lin2026alert} is the closest methodological cousin to RTV. It also tracks refusal signals across layers, but uses an internal-discrepancy amplification heuristic rather than a learned outlier score and operates in a zero-shot regime without calibration. Finally, Circuit Breakers~\cite{zou2024improving} take a training-time approach, using representation rerouting to short-circuit harmful representations before they reach the output. Across this family, each defense, with the partial exception of Alert, collapses the model's layer-wise computation into a single decision point, one gradient, one layer pair, one concept-gate, and discards the layer-wise dynamics along which jailbreak and legitimate prompts diverge. RTV differs by consuming the full layerwise refusal-direction trajectory as a multi-dimensional fingerprint and scoring it with a Mahalanobis detector, applying distance-based OOD detection to an interpretable representation-level signal.

\subsection{Out-of-Distribution Detection with Mahalanobis Distance}
\label{sec:related_mahalanobis}

Mahalanobis distance on intermediate features is a mature primitive for OOD and adversarial-example detection. Lee~\emph{et~al.}~\cite{lee2018simple} establish a framework that combines per-layer Mahalanobis scores via logistic regression and achieves strong performance on both OOD and adversarial inputs. Hendrycks~\emph{et~al.}~\cite{hendrycks2020pretrained} show that pretrained Transformers substantially improve OOD robustness over prior architectures. Podolskiy~\emph{et~al.}~\cite{podolskiy2021revisiting} demonstrate state-of-the-art OOD intent detection with Mahalanobis distance on fine-tuned Transformer embeddings.

Several recent works apply this machinery to LLM safety tasks. Representational Contrastive Scoring (RCS)~\cite{hua2025rethinking} instantiates a contrastive Mahalanobis detector for large vision-language models, scoring inputs at a single safety-critical layer by their relative distance to benign versus malicious distributions. MAAD~\cite{feng2026benchmarking}, a benchmark for LLM misalignment anomaly detection, reports that Mahalanobis distance on hidden representations is the strongest baseline overall across seven failure modes. PALE~\cite{li2026bolster} applies a contrastive Mahalanobis score to intermediate activations for hallucination detection and demonstrates that the same machinery extends beyond OOD detection to factuality monitoring. JailDAM~\cite{nian2025jaildam} pursues a complementary direction, modeling the distribution of benign inputs with an autoencoder and using reconstruction error as an anomaly score. RCS reports outperforming JailDAM by adopting a contrastive rather than one-class formulation.

All of these detectors operate on raw hidden states rather than on refusal-direction projections, and typically at a single layer and token position. RTV applies Mahalanobis scoring to a refusal-direction fingerprint computed over multiple layers and token positions. RTV retains the statistical properties of Mahalanobis-based OOD detection while operating on an attack-specific signal absent from raw activations.

\subsection{Refusal-Direction Analysis and Representation Engineering}
\label{sec:related_refusal}

A parallel line of work identifies and manipulates safety-relevant directions in the residual stream. The representation-engineering framework of Zou~\emph{et~al.}~\cite{zou2023representation} establishes that high-level model behaviors, including truthfulness, refusal, and power-seeking, can be localized to low-dimensional subspaces and controlled through targeted interventions. Contrastive Activation Addition (CAA)~\cite{panickssery2023steering} implements this by extracting steering vectors from contrastive prompt pairs and adding them to the residual stream to elicit or suppress specific behaviors. Arditi~\emph{et~al.}~\cite{arditi2024refusal} apply this methodology specifically to refusal, showing that refusal is mediated by a single direction across 13 chat models. They show that adversarial suffixes such as GCG~\cite{zou2023universal} succeed by suppressing propagation along this refusal-mediating direction.

HiddenDetect~\cite{jiang2025hiddendetect} is the closest prior instantiation of refusal signals as a detection feature. It constructs a refusal vector from high-logit refusal tokens (e.g., ``sorry'', ``unable'') in vocabulary space and scores inputs by the cosine similarity between their projected hidden states at the final token position and this refusal vector, aggregated over a set of safety-aware layers. 

\textbf{RTV differs from HiddenDetect along three axes:} First, our refusal direction is the Arditi residual-stream difference-of-means direction that GCG suppresses, not a vocabulary-space construction from refusal tokens. Second, our readout spans multiple layers and token positions rather than the final token position alone. Third, cosine similarities serve as \emph{features} for Mahalanobis outlier scoring rather than as a detection score directly, which gives it the statistical properties of a well-studied OOD framework.

\subsection{Adaptive Evaluation of Safety Defenses}
\label{sec:related_adaptive}

Adaptive evaluation has become a prerequisite for credible LLM-safety claims, following the adaptive evaluation methodology of Tramer~\emph{et~al.}~\cite{tramer2020adaptive}. Carlini~\emph{et~al.}~\cite{carlini2023are} show that aligned LLMs are vulnerable to adversarial attacks in continuous input spaces, undermining claims that alignment alone provides robustness. Andriushchenko~\emph{et~al.}~\cite{andriushchenko2024jailbreaking} demonstrate that simple adaptive jailbreaks bypass leading safety-aligned LLMs across the Llama, Claude, and GPT families, emphasizing that non-adaptive evaluation systematically overestimates robustness.

Closer to our setting, Bailey~\emph{et~al.}~\cite{bailey2026obfuscated} show that state-of-the-art representation-space defenses are systematically bypassed by gradient-based obfuscated-activation attacks that jointly optimize for a harmful completion and for activations indistinguishable from benign. Their results are relevant here because they refute a naive high-dimensionality argument. Mahalanobis scoring over raw LLM activations already operates on thousands of dimensions, yet remains bypassable. Schwinn and Geisler~\cite{schwinn2024revisiting} report a similar result against Circuit Breakers, achieving $100\%$ ASR on both released models with only three modifications to the embedding-space attack used in the original evaluation. These results define the bar for evaluating new representation-space defenses, and directly motivate our adaptive evaluation of RTV in Section~\ref{sec:rtv_adaptive}.

\section{Background}

\subsection{Greedy Coordinate Gradient (GCG)}
\label{sec:bg_gcg}

Let $f_\theta$ denote a decoder-only Transformer with vocabulary $\mathcal{V}$, $L$ layers, and hidden dimension $d$. Given an input token sequence $t = (t_1, \dots, t_n) \in \mathcal{V}^n$, the model produces residual-stream activations $h_l(i) \in \mathbb{R}^d$ at each layer $l \in \{1, \dots, L\}$ and token position $i \in \{1, \dots, n\}$, which we use throughout Sections~\ref{sec:adaptiv_attack} and~\ref{sec:rtv}.

Zou~\emph{et~al.}~\cite{zou2023universal} introduce Greedy Coordinate Gradient (GCG), a white-box optimization-based jailbreak that appends an adversarial suffix to a harmful query. We partition the input into a harmful query $q \in \mathcal{V}^{n_q}$ and an adversarial suffix $x \in \mathcal{V}^{n_x}$, with concatenation denoted $q \oplus x$. Given a target affirmative response $y \in \mathcal{V}^m$, e.g., ``Sure, here is'', GCG solves
\begin{equation}
\min_{x \in \mathcal{V}^{n_x}} \; \mathcal{L}_{\mathrm{GCG}}(x) = -\log p_\theta(y \mid q \oplus x).
\label{eq:bg_gcg_loss}
\end{equation}
The minimization is over discrete tokens, which GCG handles with a gradient-guided top-$k$ substitution procedure. At each iteration, the method (i) computes the gradient of $\mathcal{L}_{\mathrm{GCG}}$ with respect to the one-hot encoding of each suffix position, (ii) selects the top-$k$ candidate replacement tokens per position from this gradient, (iii) samples a batch of $B$ candidate substitutions from the pool, evaluates the loss of each, and (iv) commits the substitution that yields the lowest loss. This loop runs for a budget of $T$ iterations, with $T = 600$ throughout this paper (Section~\ref{sec:threat_model}). Although GCG requires white-box gradient access, Zou~\emph{et~al.} further show that optimized suffixes transfer to black-box models, making the attack a standard benchmark for LLM safety~\cite{zou2023universal}. Our adaptive attacks in Sections~\ref{sec:adaptiv_attack} and~\ref{sec:rtv} modify the loss in Eq.~\eqref{eq:bg_gcg_loss} to jointly target unsafe behavior and detector evasion while preserving the top-$k$ substitution procedure.

\begin{figure*}    
    \centering
    \includegraphics[width=0.8\linewidth]{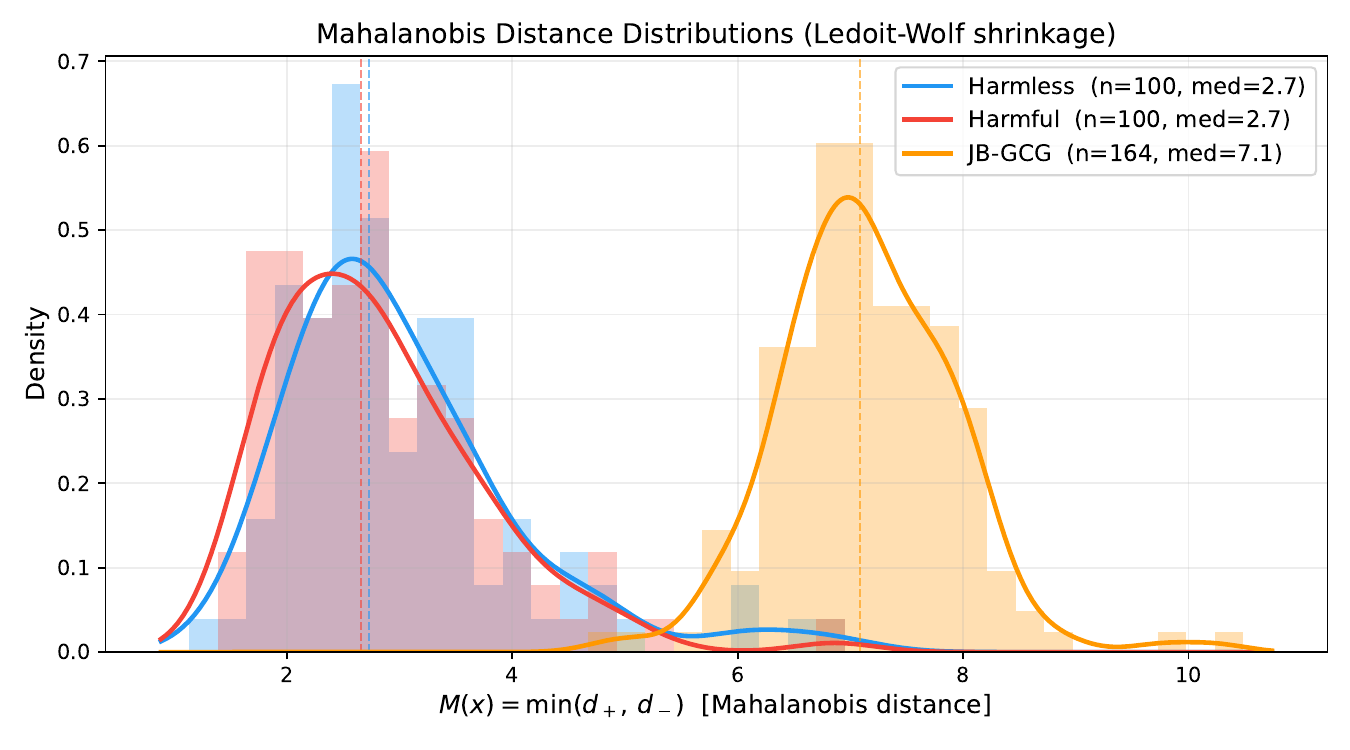}
    \caption{Mahalanobis distance distributions on the test set. The JB-GCG distribution (median 7.1) is well-separated from harmless (median 2.7) and harmful (median 2.7), with minimal overlap.}
    \label{fig:mahalanobis_histogram2}
\end{figure*}

\subsection{JBShield}
\label{sec:bg_jbshield}

JBShield~\cite{zhang2025jbshield} is a representation-level jailbreak defense comprising a detection component (JBShield-D) and a mitigation component (JBShield-M). Both operate on two concept directions extracted from last-token hidden-state activations: a \textit{toxic} direction that separates harmful from harmless prompts, and a per-attack \textit{jailbreak} direction that separates prompts of a specific jailbreak family from their underlying harmful prompts. In this paper, we introduce a GCG-style attack. Thus, we restrict our attention to the GCG jailbreak family throughout the paper.

\medskip
\noindent
\textbf{Layer selection.} Let $h_{l}(p) \in \mathbb{R}^d$ denote the final-token hidden-state activation of prompt $p$ at layer $l$. From a calibration set of $N$ harmless prompts $\{p_i^b\}$, harmful prompts $\{p_i^h\}$, and jailbreak prompts $\{p_i^{j}\}$ for GCG attack family, JBShield selects two layers. The toxic layer $l_t$ is the layer that minimizes the mean all-pairs cosine similarity between harmful and harmless calibration embeddings; a separate jailbreak layer $l_j$ is chosen for GCG attack family as the layer minimizing the mean all-pairs cosine similarity between attack embeddings and harmful embeddings. Because $l_j$ is attack-specific, JBShield-D implicitly assumes the defender knows the attack family at inference time.

\medskip
\noindent
\textbf{Concept directions.} JBShield extracts anchor concept directions by taking the top right singular vector of paired difference matrices at the selected layers. For the toxic concept, the difference matrix $\mathbf{D}_t^{\text{cal}} \in \mathbb{R}^{N \times d}$ has rows $h_{l_t}(p_i^h) - h_{l_t}(p_i^b)$ for $i = 1, \dots, N$, and
\begin{equation}
\mathbf{v}_t = \operatorname{top\text{-}SV}\!\left(\mathbf{D}_t^{\text{cal}}\right).
\label{eq:bg_anchor_toxic}
\end{equation}
The jailbreak concept direction $\mathbf{v}_j$ is analogously extracted from $\mathbf{D}_j \in \mathbb{R}^{N \times d}$, whose rows are $h_{l_j}(p_i^{j}) - h_{l_j}(p_i^h)$.

\medskip
\noindent
\textbf{JBShield-D (detection).} Given a test prompt $p$, JBShield-D computes input-specific concept directions as the unit-normalized difference between the prompt's embedding and the mean anchor of the opposing class:
\begin{equation}
\mathbf{v}_p^{\text{toxic}} = \frac{h_{l_t}(p) - \bar{h}_{l_t}^b}{\|h_{l_t}(p) - \bar{h}_{l_t}^b\|},
\qquad
\mathbf{v}_p^{\text{jailbreak}} = \frac{h_{l_j}(p) - \bar{h}_{l_j}^h}{\|h_{l_j}(p) - \bar{h}_{l_j}^h\|},
\label{eq:bg_input_vectors}
\end{equation}
where $\bar{h}_{l_t}^b$ and $\bar{h}_{l_j}^h$ are the mean harmless and harmful calibration embeddings at the respective layers. It then scores $p$ by cosine similarity against the anchor concept directions:
\begin{equation}
s_t = \cos\!\left(\mathbf{v}_p^{\text{toxic}},\, \mathbf{v}_t\right),
\qquad
s_j = \cos\!\left(\mathbf{v}_p^{\text{jailbreak}},\, \mathbf{v}_j\right).
\label{eq:bg_scores}
\end{equation}
JBShield-D forms an AND-gate and a prompt is flagged as a jailbreak only if both scores exceed thresholds $T_t$ and $T_j$:
\begin{equation}
R(p) =
\begin{cases}
\text{True}, & \text{if } s_t \geq T_t \text{ and } s_j \geq T_j,\\
\text{False}, & \text{otherwise.}
\end{cases}
\label{eq:bg_decision}
\end{equation}
This AND-gate is central to JBShield's design. A prompt evades detection if it fails either threshold.

\medskip
\noindent
\textbf{JBShield-M (mitigation).} JBShield-M conditionally adds a scaled concept vector: $h_l \leftarrow h_l + \delta_t \mathbf{v}_t$ when the toxic concept is detected, and $h_l \leftarrow h_l - \delta_j \mathbf{v}_j$ when the jailbreak concept is detected. The two operations operate independently and do not share an AND-gate. A prompt that bypasses JBShield-D can still be partially neutralized by JBShield-M if one or both detections activate; conversely, an attack that evades both concepts sees no mitigation at all. We evaluate both JBShield-D and JBShield-M in Section~\ref{sec:jb-d-m-gcg}.

\subsection{Refusal Direction}
\label{sec:bg_refusal}

Recent work shows that refusal behavior in aligned LLMs is localized to a low-dimensional subspace of the residual stream~\cite{zou2023representation, panickssery2023steering, arditi2024refusal}. Arditi~\emph{et~al.}~\cite{arditi2024refusal} identify a single \textit{refusal direction} that mediates this behavior across $13$ open-weight chat models, estimated via a difference-in-means construction over contrastive prompt sets.

Let $h_{l}(p) \in \mathbb{R}^d$ denote the final-token hidden-state activation of prompt $p$ at layer $l$, consistent with the notation in Section~\ref{sec:bg_jbshield}. Given a calibration set of harmful prompts $\mathcal{D}^h$ and harmless prompts $\mathcal{D}^b$, the refusal direction at layer $l$ is
\begin{equation}
r_l = \mu_l^{h} - \mu_l^{b},
\qquad
\mu_l^{h} = \frac{1}{|\mathcal{D}^h|}\sum_{p \in \mathcal{D}^h} h_{l}(p),
\qquad
\mu_l^{b} = \frac{1}{|\mathcal{D}^b|}\sum_{p \in \mathcal{D}^b} h_{l}(p),
\label{eq:bg_refusal_direction}
\end{equation}
with unit-normalized form $\hat{r}_l = r_l / \|r_l\|$.

Arditi~\emph{et~al.} establish that $\hat{r}_l$ is a causal mediator of refusal: adding $\hat{r}_l$ to the residual stream elicits refusal on harmless prompts, and ablating the component along $\hat{r}_l$ disables refusal on harmful prompts while preserving general capabilities. More relevant to our setting, they further show that adversarial suffixes including GCG succeed by suppressing propagation along $\hat{r}_l$ during the forward pass~\cite{arditi2024refusal}, achieving the same mechanistic effect as directional ablation via optimization over the input rather than a surgical intervention on hidden states. This makes the cosine alignment $\cos(h_l(p), \hat{r}_l)$ a mechanistically grounded attack signature: successful jailbreaks should exhibit reduced alignment relative to harmful or harmless prompts. RTV (Section~\ref{sec:rtv}) builds on this observation by treating per-layer refusal-direction cosines as features for outlier detection.

\subsection{Mahalanobis Distance for Outlier Detection}
\label{sec:bg_mahalanobis}

Mahalanobis distance is a standard scoring function for out-of-distribution and adversarial-example detection in deep feature spaces~\cite{lee2018simple, podolskiy2021revisiting}. Given a feature map $\mathbf{f}: \mathcal{X} \to \mathbb{R}^k$ and a reference distribution with mean $\boldsymbol{\mu}$ and covariance $\boldsymbol{\Sigma}$, the Mahalanobis distance of a point $x$ is
\begin{equation}
    d(x) = \sqrt{(\mathbf{f}(x) - \boldsymbol{\mu})^\top \boldsymbol{\Sigma}^{-1} (\mathbf{f}(x) - \boldsymbol{\mu})}.
    \label{eq:bg_mahalanobis}
\end{equation}
At deployment, one fits $\hat{\boldsymbol{\mu}}$ and $\hat{\boldsymbol{\Sigma}}$ on in-distribution calibration features, computes $d(x)$ for test points, and flags points exceeding a threshold as outliers. For RTV, $\mathbf{f}(x)$ is the refusal-direction fingerprint described in Section~\ref{sec:rtv}, whose entries correspond to cosine similarities at different layers and token positions. These coordinates have heterogeneous variances and strong cross-layer correlations, making Mahalanobis the natural scoring choice. It normalizes by per-coordinate variance and accounts for correlations. Because the fingerprint is low-dimensional ($k=15$) but the calibration set is also small, the sample covariance $\hat{\boldsymbol{\Sigma}}$ can be ill-conditioned. We use Ledoit-Wolf shrinkage~\cite{ledoit2004well} which yields a well-conditioned estimate. Fig.~\ref{fig:mahalanobis_histogram2} shows that the Mahalanobis distance distribution of JB-GCG, our adversarial prompts we introduce in Section~\ref{sec:jb-d-m-gcg}, is well-separated from the distribution of legitimate (harmless and harmful) prompts. Motivated by this separation, we construct a new defense that uses Mahalanobis distance as an outlier detector, which evaluates if the test prompt is a legitimate prompt. A proper defense needs to detect various families of jailbreaks, not just a single family. We show in Section~\ref{sec:rtv_results} that Mahalanobis outlier detector works well across different jailbreak types, not just JB-GCG.

\section{Attack--Defense Cycle Overview}
\label{sec:threat_model}

\begin{figure}    
    \centering
    \includegraphics[width=\linewidth]{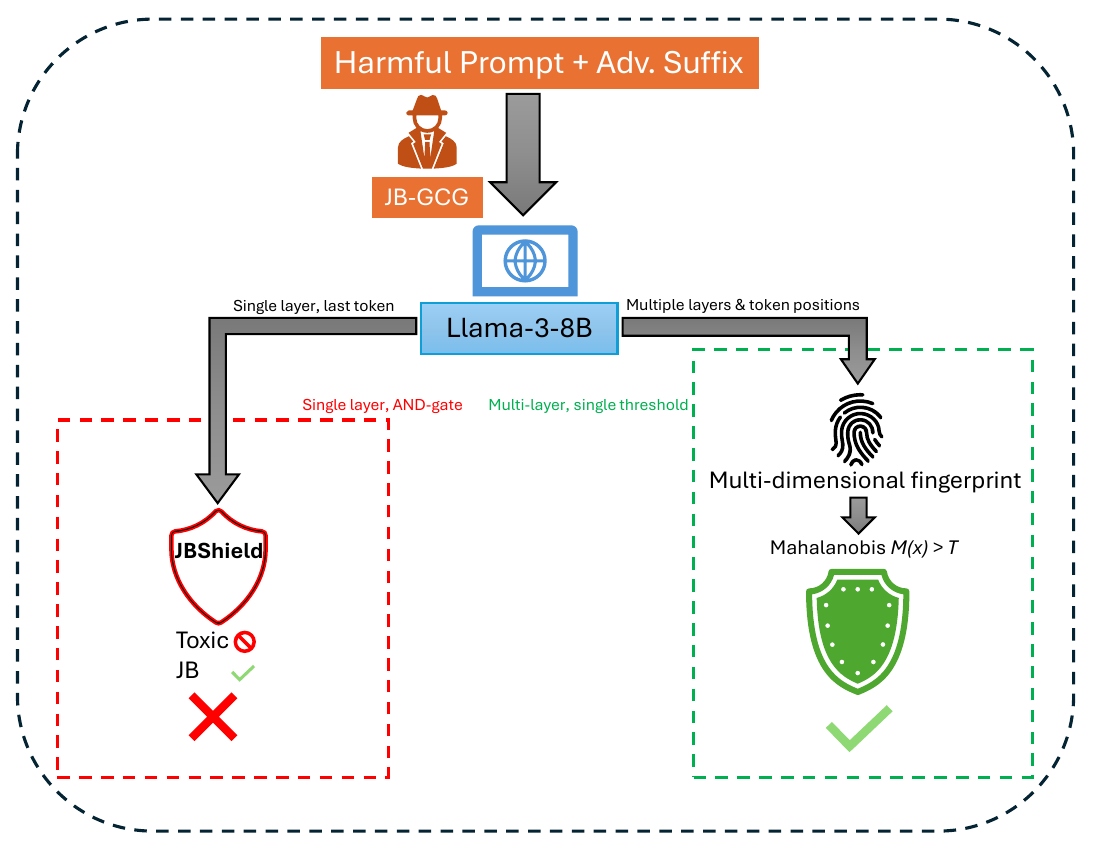}
    \caption{Overview of the attack-defense cycle. \textbf{Left:} JBShield-D extracts hidden states at a single frozen layer and checks two concept scores (toxic, jailbreak) via an AND-gate. JB-GCG suppresses the toxic concept below threshold while leaving the jailbreak concept active, breaking the detection. \textbf{Right:} RTV extracts hidden states at multiple layers and token positions, computes cosine similarities with per-layer refusal directions to form a multi-dimensional fingerprint, and flags outliers via Mahalanobis distance.}
    \label{fig:overview}
\end{figure}

Figure~\ref{fig:overview} illustrates the attack-defense cycle studied in this paper. JB-GCG bypasses JBShield's single-layer AND-gate detection (left), while RTV detects the same attack using multi-layer fingerprinting with Mahalanobis outlier scoring (right).

\subsection{Threat Model}

We consider an adaptive white-box threat model that reflects the worst case for a deployed LLM defense. Throughout this paper, we target \texttt{Meta-Llama-3-8B-Instruct}~\cite{llama3modelcard}, which we refer to as Llama-3-8B for brevity. 

\medskip
\noindent
\textbf{Attacker.} 
The attacker's goal is to induce the target aligned LLM to produce unsafe outputs on harmful prompts drawn from the HarmBench evaluation set~\cite{mazeika2024harmbench} while simultaneously evading the deployed detector. The attacker has full white-box access to the target model, including weights, activations, and gradients, as required for GCG-style suffix optimization~\cite{zou2023universal}. For adaptive attack design against JBShield in Section~\ref{sec:adaptiv_attack}, the attacker has full white-box access to JBShield defense, e.g.,  the calibration embeddings, the anchor concept directions, and the concept detection thresholds. The attacker can modify GCG's objective function to target JBShield's defense mechanism. For the adaptive evaluation of RTV in Section~\ref{sec:rtv_adaptive}, the attacker additionally has full white-box access to the defense, e.g., the refusal direction, the set of layers and token positions comprising the fingerprint, the calibrated Mahalanobis mean and covariance, and the detection threshold. The attacker cannot modify the model weights or the defender's calibration set. The attack surface is limited to injection of a token-level adversarial suffix appended to the original prompt~\cite{zou2023universal}. Attack cost is bounded by an optimization budget of 600 GCG iterations per prompt.

\medskip
\noindent
\textbf{Defender.} 
The defender has white-box access to the target model's hidden states at all layers and token positions, but does not retrain or fine-tune the base model. The defender possesses a small calibration set of harmless and harmful prompts used to estimate the refusal direction~\cite{arditi2024refusal} and the Mahalanobis parameters. No jailbreak examples are required. The defender knows that adaptive jailbreak attacks exist and may target the defense, but does not have access to the specific suffixes an attacker will use at deployment time.

\medskip
\noindent
\textbf{Success criteria.} 
We measure attack success using ASR, defined as the fraction of harmful prompts on which the attacker produces an output judged unsafe by SORRY-Bench~\cite{xie2024sorry} and not flagged by the detector. We report this metric separately against JBShield-D, JBShield-M, and RTV.

\section{Adaptive Attack Design} \label{sec:adaptiv_attack}

JBShield~\cite{zhang2025jbshield} reports 0\% ASR against GCG-based adaptive attacks in which the attack objective is modified to weaken the toxic concept and enhance the jailbreak concept (Table~9 in~\cite{zhang2025jbshield}). However, the specific loss formulation used for this adaptive evaluation is not described in detail, and no implementation is publicly available at the time of writing. In our own experiments, standard GCG with target-string optimization does not produce adversarial suffixes that evade JBShield. Motivated by this, we develop an adaptive attack that replaces the target-string objective with representation-level optimization terms that evade JBShield's detection.

Our goal is to generate adversarial suffixes that satisfy two conditions simultaneously: they should weaken the base model's refusal behavior enough to induce an unsafe response, and they should evade JBShield detection. We first explore a refusal-direction based objective that directly weakens refusal in activation space. This objective can reduce refusal behavior, but it does not reliably suppress JBShield's detection on the \textit{toxic} and \textit{jailbreak} concepts. We then introduce detector-aware objectives that directly target JBShield's two concept scores in a joint manner. Finally, we combine refusal suppression with JBShield-aware regularization to obtain our final method, \textbf{JB-GCG}.

\subsection{Directional-GCG}

We begin with an objective that directly suppresses refusal behavior in activation space. Let $\mathbf{r}_{l}$ denote the refusal direction at layer~$l$ extracted as shown in Section~\ref{sec:bg_refusal}. Let $\mathbf{h}_{l}$ denote the hidden-state representation at the final token position when the harmful query, $q$, is appended by an adversarial suffix, $x$. If refusal behavior is encoded along $\mathbf{r}_{l}$, then the cosine similarity between $\mathbf{r}_{l}$ and $\mathbf{h}_{l}$ measures how strongly the current hidden-state representation aligns with refusal-related features. Based on this idea, we define \textbf{Directional-GCG} as
\begin{equation}
\min_{x \in \mathcal{V}^{n_x}} \; \mathcal{L}_{\mathrm{D\text{-}GCG}}(x)
=
-\cos(\mathbf{r}_{l},\, \mathbf{h}_{l}).
\label{eq:dgcg_loss}
\end{equation}
This objective pushes the hidden representation away from the refusal direction and therefore targets refusal more directly. Compared with target-string optimization in standard GCG, this objective is also less restrictive, since it does not require the model to match a fixed affirmative response. Instead, it searches for adversarial suffixes that broadly reduce refusal-related alignment in activation space. Directional-GCG can produce suffixes that suppress refusal behavior. However, because it does not explicitly optimize against JBShield's detector, prompts generated by this method may still be flagged as jailbreak attempts.

\subsection{Joint-GCG}

The previous objective targets refusal behavior, but refusal evasion alone is insufficient if JBShield still detects the harmful prompt. We therefore introduce a detector-aware baseline that directly optimizes against JBShield's two concept scores. Let $\mathbf{v}^{\text{jailbreak}}$ and $\mathbf{v}^{\text{toxic}}$ denote the input-dependent concept representations extracted by JBShield with the suffix, $x$, and let $\mathbf{v}_j$ and $\mathbf{v}_t$ denote the corresponding anchor concept representations. We define \textbf{Joint-GCG} as
\begin{equation}
\min_{x \in \mathcal{V}^{n_x}} \; \mathcal{L}_{\mathrm{J\text{-}GCG}}(x)
=
-
\alpha \cos(\mathbf{v}^{\text{toxic}},\, \mathbf{v}_t)
-
\beta \cos(\mathbf{v}^{\text{jailbreak}},\, \mathbf{v}_j)
,
\label{eq:jgcg_loss}
\end{equation}
where $\alpha$ and $\beta$ control the importance of the jailbreak and toxic concept terms. This objective serves as a detector-evasion baseline: it tests whether directly minimizing JBShield's two concept-alignment scores is sufficient to produce adversarial suffixes that bypass the detector. Since JBShield requires both concept scores to exceed threshold before flagging an input, reducing either score may be sufficient for evasion. However, jointly minimizing both terms can introduce optimization interference, and detector-only optimization might not suppress the model's refusal behavior.

\subsection{JB-GCG}

The above observations suggest that a successful attack against JBShield must satisfy two conditions simultaneously: it must suppress refusal in the base model and reduce detector-side concept activation sufficiently to evade detection. Motivated by this, we define our final method, \textbf{JB-GCG}, by combining the Directional-GCG objective with a JBShield-aware regularization term. 
The choice of which concept to regularize depends on the concept which yields higher ASR. The experiment in Section~\ref{sec:joint_gcg} shows that the toxic concept is the dominant signal under JBShield's detection, which motivates the toxic-concept regularizer in the final objective.
\begin{equation}
\min_{x \in \mathcal{V}^{n_x}} \; \mathcal{L}_{\mathrm{JB\text{-}GCG}}(x)
=
-\cos(\mathbf{r}_{l},\, \mathbf{h}_{l})
+
\cos^2(\mathbf{v}^{\text{toxic}},\, \mathbf{v}_t).
\label{eq:jbgcg_loss}
\end{equation}
The first term suppresses refusal by reducing alignment with the refusal direction, while the second term penalizes alignment with JBShield's toxic concept representation. We use the squared cosine in Eq.~\eqref{eq:jbgcg_loss} rather than the linear form because its gradient vanishes as the toxic concept score approaches zero, allowing the optimizer to shift focus to refusal suppression once evasion is achieved. These terms encourage adversarial suffixes that both weaken refusal and evade representation-level detection. Empirically, this objective consistently suppresses the toxic concept while leaving the jailbreak concept active, which is sufficient to evade JBShield because its detector requires both concepts to exceed threshold. We use JB-GCG as the final attack method in our evaluation.

\subsection{Optimization Details}

In all experiments, we define $(l_{\text{dir}}, l_{\text{opt}})$ as the layer from which the refusal direction is extracted and the layer at which the hidden-state representations are optimized, respectively. We extract hidden-state representations at the final token position for all cosine-similarity computations. For detector-aware objectives, optimization terminates once the current prompt is no longer flagged by JBShield. For JB-GCG, this means that optimization stops when the toxic concept cosine similarity falls below its corresponding threshold and the generated response is non-refusal. If these conditions are not satisfied, optimization continues until the iteration budget is exhausted. Unless otherwise stated, all experiments are conducted on Llama-3 8B with 100 randomly sampled HarmBench prompts. 

\section{Experiments}

We first start by evaluating whether refusal suppression is effective enough to break JBShield-D. Then, we evaluate whether detector-aware optimization alone is sufficient to evade JBShield-D. In our last attack, we evaluate whether combining refusal suppression with detector-aware regularization yields successful jailbreaks against the full JBShield defense pipeline, including JBShield-M. We separately report toxic and jailbreak concept detections in order to determine which part of JBShield's decision rule is being broken by the attack. ASR against both JBShield-D and JBShield-M are evaluated by SORRY-Bench~\cite{xie2024sorry}. \textbf{Generated / Total} denotes the number of prompts for which optimization produced a candidate suffix. \textbf{Toxic Detection} and \textbf{JB Detection} report the percentage of generated suffixes detected by the toxic and jailbreak concepts, respectively. \textbf{ASR} reports the percentage of generated suffixes that both induce harmful behavior and evade JBShield.

\subsection{Directional-GCG against JBShield}

\begin{table}
\caption{Directional-GCG results for different refusal-direction and optimization-layer settings. ASR shows the percentage of successful attacks that both induce harmful output from the model and evade JBShield-D.}
\label{tab:dgcc_results}
\centering
\footnotesize
\setlength{\tabcolsep}{3pt}
\begin{tabular}{ccccccc}
\toprule
\textbf{$l_{\text{dir}}$} & \textbf{$l_{\text{opt}}$} & \textbf{Dataset} & \textbf{Generated / Total} & \textbf{Toxic Detect.} & \textbf{JB Detect.} & \textbf{ASR} \\
\midrule
18 & 18 & HarmBench     & 98 / 100 & 92.8 & 100.0 & 6.1 \\
18 & 18 & JBShield test & 99 / 100 & 95.6 & 100.0 & 1.0 \\
17 & 32 & HarmBench     & 95 / 100 & 97.8 & 100.0 & 1.0 \\
18 & 32 & HarmBench     & 99 / 100 & 96.9 & 100.0 & 2.0 \\
26 & 32 & HarmBench     & 97 / 100 & 96.9 & 100.0 & 3.0 \\
\bottomrule
\end{tabular}

\end{table}

Before constructing the attack, we must select the layer from which to extract the refusal direction. We perform a layer-wise ablation study on Llama-3-8B, ablating the refusal direction at each layer and measuring the resulting safety score (Appendix~\ref{app:chat_ablation}). Layers 17--22 yield the lowest safety scores after ablation, indicating that refusal behavior is most strongly expressed in this range. We select layer~18 as the primary extraction layer for Directional-GCG. Table~\ref{tab:dgcc_results} reports results across several refusal-direction extraction layer, optimization layer, and dataset configurations. In the primary configuration, $(l_{\text{dir}}, l_{\text{opt}})=(18,18)$ on HarmBench, Directional-GCG successfully generates candidate suffixes for 98 out of 100 prompts, confirming that refusal-direction optimization effectively suppresses the model's refusal behavior. However, JBShield detects nearly all generated suffixes: the toxic concept detection flags 92.8\% (91 suffixes) and the jailbreak concept detector on 100\% (98 suffixes). As a result, only 6 suffixes both evade detection and elicit harmful output, yielding an ASR of 6.1\%. To assess whether the choice of dataset affects this pattern, we evaluate the same layer configuration on the JBShield test set. 99 out of 100 suffixes are generated, with toxic and jailbreak detection rates of 95.6\% and 100\%, respectively, and an ASR of only 1\%. This confirms that the low attack effectiveness is not an artifact of the evaluation dataset.

We further investigate whether varying the optimization layer relative to the direction extraction layer changes this outcome. Decoupling the two layers, $(l_{\text{dir}}, l_{\text{opt}})=(18,32)$, produces 99 suffixes but yields toxic and jailbreak detection rates of 96.9\% and 100\%, respectively, with an ASR of only 2\%. Shifting the direction extraction to a lower layer $(l_{\text{dir}}, l_{\text{opt}})=(17,32)$ similarly results in near-total detection and an ASR of 1\%. Extracting the refusal direction from a higher layer $(l_{\text{dir}}, l_{\text{opt}})=(26,32)$ likewise fails to improve evasion, achieving an ASR of 3.1\%. Across all configurations, the jailbreak concept detection flags 100\% of generated suffixes, and the toxic concept detection consistently exceeds 92\%. These results confirm that refusal-direction optimization systematically drives the model's hidden states toward the concept subspaces that JBShield monitors, regardless of the layer choices involved. This demonstrates that refusal suppression alone is fundamentally insufficient against JBShield, and motivates the addition of an explicit detector-aware regularization term.

\subsection{Joint-GCG against JBShield} \label{sec:joint_gcg}

We investigate whether JBShield can be evaded by directly modifying GCG's objective to optimize detector-side concept similarities. Figure~\ref{fig:alternating_gcg_combined} (Appendix B) shows cosine-similarity plot of toxic and jailbreak concepts for four Joint-GCG settings. When only the toxic concept is targeted ($\alpha{=}1.0, \beta{=}0.0$), the toxic score decreases substantially but remains above its detection threshold, while the jailbreak score rises sharply (Figure~\ref{fig:joint_toxic_only}). This inverse coupling reveals that the two concept signals are tightly linked under token-level optimization: token updates that reduce one score tend to elevate the other. When only the jailbreak concept is targeted ($\alpha{=}0.0, \beta{=}1.0$), optimization terminates within a few steps (Figure~\ref{fig:joint_jb_only}), indicating that jailbreak concept evasion is trivially achievable. However, no harmful outputs are produced, confirming that jailbreak detection evasion alone is insufficient. We present the cosine-similarity trajectories of JBShield concepts under two different Joint-GCG weights in Appendix~\ref{app:joint_gcg_intermediate}.

\begin{table}
\caption{Joint-GCG results for different weight settings. $\alpha$ and $\beta$ values are used to tune the weights of concept similarities.}
\label{tab:joint_results}
\centering
\footnotesize
\setlength{\tabcolsep}{3pt}
\begin{tabular}{cccccc}
\toprule
\textbf{$\alpha$} & \textbf{$\beta$} & \textbf{Generated / Total} & \textbf{Toxic Detect.} & \textbf{JB Detect.} & \textbf{ASR} \\
\midrule
0.0 & 1.0      & 100 / 100 & 100.0\% & 0.0\% & 0.0\% \\
0.5 & 1.0      & 100 / 100 & 92.0\% & 8.0\% & 6.0\% \\
0.5 & 0.5      & 99 / 100 & 72.7\% & 27.2\% & 13.1\% \\
\textbf{1.0} & \textbf{0.0} & \textbf{94 / 100} & \textbf{0.0\%} & \textbf{100.0\%} & \textbf{50.0\%} \\
1.0 & 0.5      & 82 / 100 & 1.2\% & 100.0\% & 40.2\% \\
1.0 & 1.0      & 98 / 100 & 72.4\% & 27.5\% & 14.2\% \\
\bottomrule
\end{tabular}
\end{table}

Table~\ref{tab:joint_results} confirms two patterns. First, targeting the toxic concept ($\alpha \geq 1.0$) yields substantially higher ASR than targeting the jailbreak concept, reaching 50\% when only the toxic concept is optimized which suggests that toxic concept evasion is the more attack-relevant objective. Second, at 50\% ASR, nearly half of the evading suffixes fail to elicit harmful outputs, showing that detector evasion alone might under-perform without explicit refusal suppression. This motivates the combined objective in Section~\ref{sec:jb-gcg}.

\subsection{JB-GCG against JBShield} \label{sec:jb-d-m-gcg}

\medskip
\noindent
\textbf{JB-GCG against JBShield-D.} \label{sec:jb-gcg}
We next evaluate JB-GCG across different refusal-direction and optimization-layer settings. Table~\ref{tab:layer_transfer_results} reports the number of prompts for which optimization generated a candidate suffix, the toxic and jailbreak detection rates among those generated suffixes, and the final ASR. Table~\ref{tab:layer_transfer_results} shows that JB-GCG improves over the detector-only baselines and over refusal suppression without detector-aware regularization. The strongest setting is $(l_{\text{dir}}, l_{\text{opt}})=(18,18)$ on HarmBench, where JB-GCG produces candidate suffixes for 88 out of 100 prompts and achieves an ASR of 53.4\% with zero toxic detections. On the JBShield test set, the same layer setting remains effective, producing 95 candidate suffixes out of 100 prompts and achieving an ASR of 48.4\%. For optimization at layer~32, the best result is obtained with $(26,32)$, which achieves 46.8\% ASR with zero toxic detections. Even when using a refusal direction extracted from layer~17 and applied at layer~32, JBShield detects only 1 of 45 generated suffixes through the toxic concept while the ASR still reaches 40.6\%.

\begin{table*}
\caption{JB-GCG results for different refusal-direction and optimization-layer settings. ASR-D and ASR-M denote attack success rate against JBShield-D and JBShield-M, respectively. The second row uses the JBShield test set; all other rows use HarmBench.}
\label{tab:layer_transfer_results}
\centering

\begin{tabular}{cccccccc}
\toprule
\textbf{$l_{\text{dir}}$} & \textbf{$l_{\text{opt}}$} & \textbf{Dataset} & \textbf{Generated / Total} & \textbf{Toxic Detect.} & \textbf{JB Detect.} & \textbf{ASR-D} &\textbf{ASR-M} \\
\midrule
\textbf{18} & \textbf{18} & \textbf{HarmBench} & \textbf{88 / 100} & \textbf{0.0\%} & \textbf{100.0\%} & \textbf{53.4\%} & \textbf{30.7\%} \\
18 & 18 & JBShield test & 95 / 100 & 0.0\% & 100.0\% & 48.4\% & 26.3\% \\
17 & 32 & HarmBench     & 91 / 100 & 1.1\% & 100.0\% & 40.6\% & 27.9\% \\
18 & 32 & HarmBench     & 93 / 100 & 0.0\% & 100.0\% & 41.9\% & 23.5\% \\
26 & 32 & HarmBench     & 96 / 100 & 1.0\% & 100.0\% & 46.8\% & 28.9\% \\
\bottomrule
\end{tabular}
\end{table*}


Notably, the jailbreak concept remains active in all final JB-GCG settings: the jailbreak detection rate is 100\% for every generated suffix in Table~\ref{tab:layer_transfer_results}. This shows that JB-GCG does not evade JBShield by suppressing both detector-side concepts. Instead, it consistently breaks the conjunction in JBShield's detection rule by reducing the toxic concept below threshold while leaving the jailbreak concept active. In other words, the final attack succeeds by exploiting the fact that JBShield requires both toxic and jailbreak concepts to be detected simultaneously. Overall, these results show that adaptive jailbreaks against JBShield require both refusal suppression and detector-aware optimization. Detector-only objectives can evade JBShield, while refusal-only objectives remain highly detectable. In contrast, JB-GCG combines both ingredients and achieves the strongest attack success rates across the evaluated settings.

\medskip
\noindent
\textbf{JB-GCG against JBShield-M.}
We further evaluate whether JB-GCG remains effective against the full JBShield defense pipeline, including the mitigation component JBShield-M. Recall that JBShield-M strengthens the toxic concept and weakens the jailbreak concept for prompts detected as jailbreaks, with the goal of steering the model toward safe behavior rather than issuing a fixed refusal output. The ASR-M column of Table~\ref{tab:layer_transfer_results} reports the attack success rate of JB-GCG against JBShield-M across different refusal-direction and optimization-layer settings.


JB-GCG remains effective even when JBShield's mitigation component is applied. The strongest result is obtained for $(l_{\text{dir}}, l_{\text{opt}})=(18,18)$ on HarmBench, where the attack reaches 30.7\% ASR. On the JBShield test set, the same setting achieves 26.3\% ASR. For optimization at layer~32, ASR remains between 23.5\% and 28.9\% across the evaluated layer settings. These results show that the attack does not only evade the detector in some cases, but can also remain effective against JBShield's hidden-state mitigation mechanism. Compared with the JBShield-D results, the ASR under JBShield-M is lower, which is expected because mitigation introduces an additional defense stage after detection. Nevertheless, the remaining attack success rates indicate that JBShield's mitigation mechanism is not sufficient to eliminate adaptive jailbreaks crafted by JB-GCG.

\medskip
\noindent
\textbf{Calibration-Size Sensitivity.}
We further study whether JB-GCG remains effective when JBShield is recalibrated with different calibration sizes. Since the calibration size $N$ changes the anchor vectors and thresholds used by JBShield, we re-run JB-GCG against each recalibrated defense instance. Table~\ref{tab:calibration_size_sensitivity} reports results for the $(l_{\text{dir}}, l_{\text{opt}})=(18,18)$ setting on HarmBench. JB-GCG remains effective across all evaluated calibration sizes. For every value of $N$, the attack generates candidate suffixes for most prompts, the jailbreak concept remains active for all generated suffixes, and the toxic concept is suppressed to zero detection. This indicates that the attack continues to break JBShield's detection rule by selectively reducing the toxic concept while leaving the jailbreak concept active. The strongest recalibrated setting is $N=40$, where JB-GCG generates suffixes for 92 out of 100 prompts and achieves 64.1\% ASR against JBShield-D and 38.0\% ASR against JBShield-M. Even for the least favorable setting, $N=20$, the attack still reaches 43.6\% ASR against JBShield-D and 26.4\% ASR against JBShield-M. These results show that JB-GCG is not tied to a single fragile calibration instance, but remains effective across multiple JBShield recalibrations.

\begin{table}
\caption{Calibration-size sensitivity of JB-GCG against JBShield for the $(l_{\text{dir}}, l_{\text{opt}})=(18,18)$ setting on HarmBench. ASR-D and ASR-M denote attack success rate against JBShield-D and JBShield-M, respectively.}
\label{tab:calibration_size_sensitivity}
\centering
\setlength{\tabcolsep}{3pt}
\begin{tabular}{cccccc}
\toprule
\textbf{$N$} & \textbf{Gen. / Total} & \textbf{Toxic Detect.} & \textbf{JB Detect.} & \textbf{ASR-D} & \textbf{ASR-M} \\
\midrule
10 & 88 / 100 & 0.0\% & 100.0\% & 59.0\% & 32.9\% \\
20 & 87 / 100 & 0.0\% & 100.0\% & 43.6\% & 26.4\% \\
30 & 88 / 100 & 0.0\% & 100.0\% & 53.4\% & 30.7\% \\
40 & 92 / 100 & 0.0\% & 100.0\% & 64.1\% & 38.0\% \\
50 & 89 / 100 & 0.0\% & 100.0\% & 53.9\% & 32.5\% \\
\bottomrule
\end{tabular}
\end{table}

\section{Representation Trajectory Verification (RTV)} \label{sec:rtv}

 
\subsection{Representation-Level Signatures of JB-GCG}
\label{sec:rtv_motivation}
 
The success of JB-GCG against JBShield raises the question of whether the attack leaves detectable traces that a better-designed defense could exploit. JBShield monitors a single concept at a single layer via rank-1 SVD, and JB-GCG succeeds by targeting exactly that layer and concept. However, the model processes representations through a number of transformer layers, and the attack's manipulation at the detection layer may produce detectable artifacts at other layers. To investigate, we extract the refusal direction ${r}_l$ at layers~18 and~32 using difference-in-means over 100 harmful and 100 harmless prompts, and compute the cosine similarity between ${r}_l$ and the hidden states of harmless, harmful, and JB-GCG prompts at multiple token positions.

\begin{figure}
    \centering
    \begin{subfigure}{\linewidth}
        \centering
        \includegraphics[width=\linewidth]{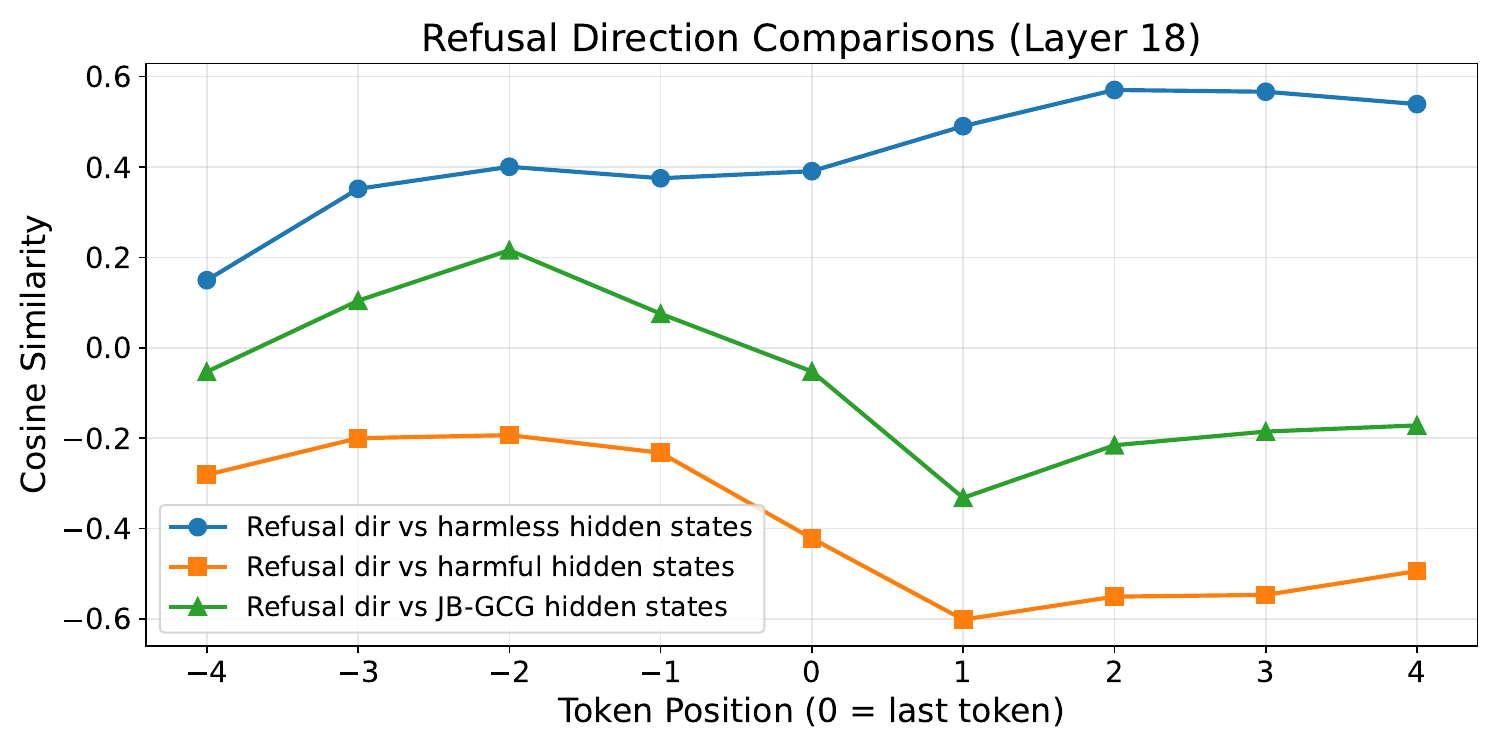}
        \caption{Cosine similarities at layer 18.}
        \label{fig:cos-layer18}
    \end{subfigure}
    \hfill
    \begin{subfigure}{\linewidth}
        \centering
        \includegraphics[width=\linewidth]{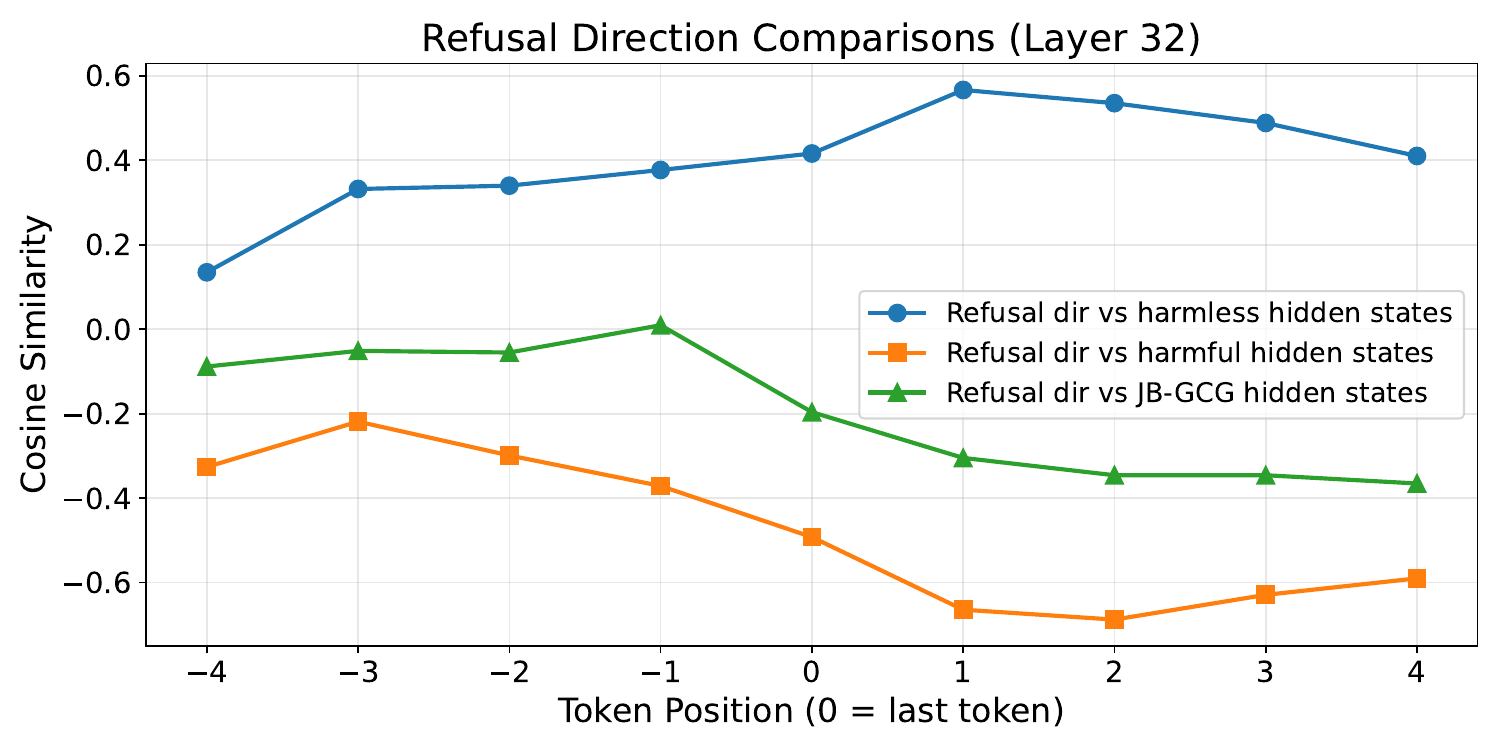}
        \caption{Cosine similarities at layer 32.}
        \label{fig:cos-layer32}
    \end{subfigure}
    \caption{Refusal-direction cosine similarities across token positions at layers 18 and 32 for harmless, harmful, and JB-GCG prompts. Legitimate categories maintain consistent sign; JB-GCG occupies a distinctive zone at layer 18 that amplifies toward harmful at layer 32.}
    \label{fig:cos-layers3_combined}
\end{figure}

\medskip
\noindent
\textbf{Layer-wise behavior.}
Figure~\ref{fig:cos-layers3_combined} shows that harmless prompts maintain consistently positive refusal-direction alignment across both layers, while harmful prompts maintain consistently negative alignment. JB-GCG prompts exhibit a qualitatively different pattern. At layer~18, the optimization target, they occupy an intermediate zone near zero. By layer~32, the model's own processing amplifies the adversarial features, pulling JB-GCG representations toward the harmful region. This \textit{inter-layer inconsistency} is absent from both legitimate categories, whose refusal-direction alignment remains stable across layers.

\begin{figure}
    \centering
    \includegraphics[width=\linewidth]{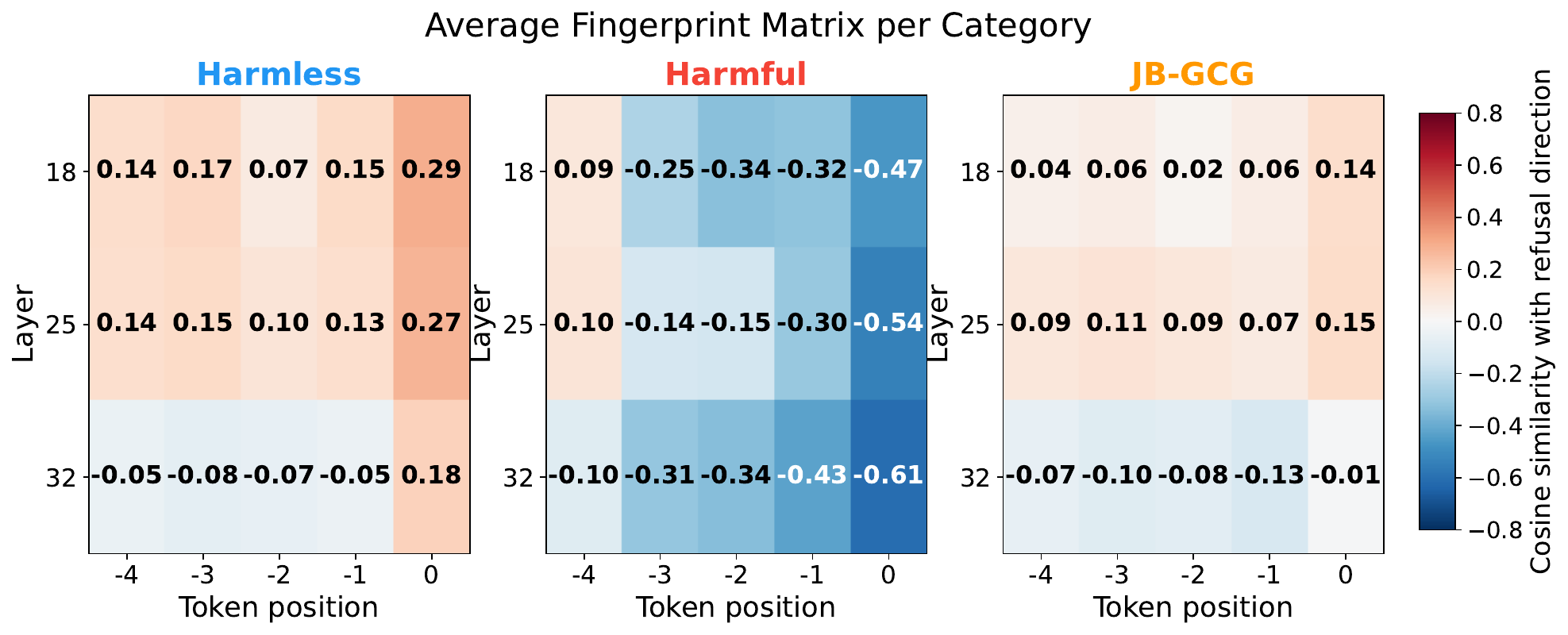}
    \caption{Average fingerprint matrix per category (3 layers $\times$ 5 positions). JB-GCG occupies a near-zero distinctive zone between the positive harmless and negative harmful profiles.}
    \label{fig:fingerprint_heatmap}
\end{figure}

\medskip
\noindent
\textbf{Fingerprint heatmap.}
Figure~\ref{fig:fingerprint_heatmap} presents the average fingerprint matrix $F \in \mathbb{R}^{3 \times 5}$ for each category across three layers (18, 25, 32) and the last five token positions. Harmless prompts are uniformly mildly positive at layers~18 and~25, with slightly negative values at layer~32 for earlier positions. Harmful prompts are strongly negative throughout, with magnitude increasing at deeper layers and later positions. JB-GCG prompts occupy a narrow band near zero, a \textit{distinctive zone} that neither legitimate category densely populates. JB-GCG's optimization pushes representations away from the refusal direction at layer~18 but cannot control how the model processes hidden-states at deeper layers, resulting in a fingerprint that matches neither harmless nor harmful profiles. Furthermore, in Appendix~\ref{app:clusters}, we show PCA and t-SNE projections of the 15-dimensional fingerprint space that confirm the cluster separations.

\subsection{Defense Design}
\label{sec:rtv_design}

RTV detects jailbreak prompts by measuring whether their refusal-direction alignment profile across layers and token positions is consistent with either legitimate category (harmless or harmful). The design is motivated by three specific JBShield vulnerabilities that JB-GCG exploits: (1) single-layer extraction (JBShield monitors one layer per concept), (2) rank-1 concept subspaces (easily targeted by gradient optimization), (3) and an AND-gate (the attacker needs to suppress only one concept). RTV avoids all three.

\medskip
\noindent
\textbf{Fingerprint construction.}
For an input prompt $x$, we extract a fingerprint matrix $F(x) \in \mathbb{R}^{K \times P}$:
\begin{equation}
    F_{l,p}(x) = \cos({r}_l,\; \mathbf{h}_l^{(p)}(x)),
    \label{eq:fingerprint}
\end{equation}
where ${r}_l$ is the normalized refusal direction at layer $l$, extracted via difference-in-means~\cite{arditi2024refusal} over harmful and harmless calibration prompts, and $\mathbf{h}_l^{(p)}(x)$ is the hidden state at layer $l$ and token position $p$. We use $K=3$ layers ($l \in \{18, 25, 32\}$) and $P=5$ token positions (the last five tokens), yielding a 15-dimensional fingerprint that is flattened to $\mathbf{f}(x) \in \mathbb{R}^{15}$. The layers are selected to span the range where refusal behavior is expressed. Layer~18 is where refusal is most sensitive. (Appendix~\ref{app:chat_ablation}), layer~32 is where the model amplifies adversarial features, and layer~25 provides intermediate coverage. Multiple token positions capture the causal-attention structure. The adversarial suffix directly influences later positions but affects earlier positions only indirectly through attention, creating a position-dependent signal that aids detection.

\medskip
\noindent
\textbf{Calibration.}
From a calibration set of $N$ harmless and $N$ harmful prompts, we compute the mean fingerprint $\boldsymbol{\mu}_+$ and covariance $\boldsymbol{\Sigma}_+$ for harmless prompts, and $\boldsymbol{\mu}_-$ and $\boldsymbol{\Sigma}_-$ for harmful prompts. With $d=15$ dimensions and $N=100$ samples per class ($N/d \approx 6.7$), the sample covariance can be ill-conditioned. We use Ledoit-Wolf shrinkage~\cite{ledoit2004well}, yielding a well-conditioned estimate without cross-validation. No jailbreak examples are required for calibration.

\medskip
\noindent
\textbf{Detection.}
For a test prompt $x$, we compute the Mahalanobis distance~\cite{lee2018simple} to both calibration clusters:
\begin{equation}
    d_+(x) = \sqrt{(\mathbf{f}(x) - \boldsymbol{\mu}_+)^\top
    \boldsymbol{\Sigma}_+^{-1} (\mathbf{f}(x) -
    \boldsymbol{\mu}_+)},
    \label{eq:mahalanobis}
\end{equation}
and analogously for $d_-(x)$. The detection score is
\begin{equation}
    M(x) = \min(d_+(x),\, d_-(x)).
    \label{eq:anomaly_score}
\end{equation}
A prompt is flagged as adversarial when $M(x) > T$, where $T$ is calibrated at the 95th percentile of the combined harmless and harmful calibration scores (targeting 5\% FPR). Our 15-dimensional fingerprint space, bounded in $[-1,1]$ and composed of cosine similarities, satisfies OOD assumptions better than the high-dimensional raw activation spaces where the method was originally validated~\cite{lee2018simple}, as confirmed by the ellipsoidal cluster structure in Figure~\ref{fig:sca-pca}.

\medskip
\noindent
\textbf{Comparison with JBShield.}
Table~\ref{tab:design_comparison} summarizes the design differences. RTV operates during a single forward pass with negligible computational overhead. Unlike JBShield's AND-gate, which an attacker can bypass by suppressing a single concept arm, RTV employs a single unified threshold over a compact fingerprint space. RTV also requires no jailbreak prompts for calibration.

\begin{table}
\caption{Design comparison between JBShield-D and RTV.}
\label{tab:design_comparison}
\centering
\footnotesize
\setlength{\tabcolsep}{3pt}
\begin{tabular}{lcc}
\toprule
\textbf{Property} & \textbf{JBShield-D} & \textbf{RTV} \\
\midrule
Layers monitored        & 1 per concept      & 3 simultaneously \\
Token positions         & Last token only    & Last 5 tokens \\
Feature extraction      & Rank-1 SVD         & Cosine similarity \\
Decision rule           & AND-gate (2 thresholds) & Single Mahalanobis threshold \\
Jailbreak examples needed & Yes ($N=30$)    & No \\
Calibration data        & $\sim$90 prompts   & $\geq$40 prompts \\
\bottomrule
\end{tabular}
\end{table}

 
\subsection{Detection Results}
\label{sec:rtv_results}

\medskip
\noindent
\textbf{Detection of JB-GCG.} 
We evaluate RTV on a test set of 100 harmless prompts from Alpaca~\cite{alpaca}, 100 harmful prompts from HarmBench~\cite{mazeika2024harmbench}, and 164 JB-GCG adversarial prompts (all prompts generated by different JB-GCG configurations from Section~\ref{sec:jb-gcg}). From the same datasets, the calibration set consists of a separate 100 harmless and 100 harmful prompts used for refusal-direction extraction and Mahalanobis parameter estimation. No jailbreak prompts are used during calibration. Figure~\ref{fig:mahalanobis_histogram2} shows the distribution of Mahalanobis scores $M(x)$ across the three categories. The JB-GCG distribution (median 7.1) is well-separated from both the harmless (median 2.7) and harmful (median 2.7) distributions, with minimal overlap. Table~\ref{tab:rtv_headline} reports the headline detection metrics. At the 5\% FPR operating point, RTV detects all of 164 JB-GCG prompts, the same attack that achieves 53.4\% ASR against JBShield-D, which detects 0\% of these prompts via its toxic concept. Furthermore, we verify that RTV’s performance is not sensitive to the specific calibration data in Appendix~\ref{app:calibration_stability2}, showing that AUROC varies by only 0.003 across splits, while TPR remains at or above 99.4\%. In Appendix~\ref{app:calibration_size}, we evaluate RTV across calibration sizes $N \in \{10, 20, 30, 50, 100\}$ per class, showing that RTV requires as few as $N=20$ calibration samples per class to achieve $\geq$97\% detection. In Appendix~\ref{app:ablations}, we present our ablation experiments on RTV, showing that using 3-layer configuration offers 100\% detection against JB-GCG and using the last five tokens improves the detection performance compared to using the last three token.
 
\begin{table}
\caption{RTV detection performance against JB-GCG.}
\label{tab:rtv_headline}
\centering
\begin{tabular}{lc}
\toprule
\textbf{Metric} & \textbf{Value} \\
\midrule
AUROC & 0.9946 \\
TPR @ 5\% FPR & 100.00\% (164/164) \\
TPR @ 1\% FPR & 81.10\% (133/164) \\
\bottomrule
\end{tabular}
\end{table}

\medskip
\noindent
\textbf{Generalization Across Attack Types.} 
To evaluate whether RTV generalizes beyond JB-GCG, we test it on nine standard jailbreak attacks from the JBShield evaluation suite, using 50 prompts per attack on Llama-3-8B. The same calibration statistics are used without recalibration. Table~\ref{tab:rtv_generalization} shows that RTV achieves perfect detection on five of nine attacks (SAA, DrAttack, Puzzler, Zulu, Base64) and near-perfect on IJP (0.96), matching or exceeding JBShield-D. RTV underperforms JBShield-D on AutoDAN (0.72 vs.\ 0.97) and GCG (0.88 vs.\ 0.98), and both defenses struggle with PAIR (0.58 vs.\ 0.77). PAIR generates naturalistic prompts via iterative LLM refinement, producing representations that are inherently close to the harmless distribution, a fundamental challenge for any representation-level detector.  The gap between RTV and JBShield-D on standard attacks reflects a fundamental design tradeoff. JBShield-D calibrates a separate jailbreak layer and concept direction per attack family, requiring 30 jailbreak examples from each target attack type. JBShield-D assumes the defender knows the attack family at the inference time. RTV requires no jailbreak examples and uses a single fixed detector across all attacks, without assuming the attack family. Under this zero-shot constraint, RTV underperforms JBShield-D on attacks whose representations are close to the harmless distribution (PAIR, AutoDAN) but matches or exceeds it on six others. The critical difference is adaptive robustness. JBShield-D achieves 0\% detection against JB-GCG, while RTV detects 100\% of the same attack using the same calibration. 
 
\begin{table}
\caption{RTV detection accuracy across standard jailbreak attacks on Llama-3-8B, compared with JBShield-D. RTV uses Mahalanobis based outlier detector at 5\% FPR.}
\label{tab:rtv_generalization}
\centering
\begin{tabular}{lccc}
\toprule
\textbf{Attack} & \textbf{$n$} & \textbf{JBShield-D} & \textbf{RTV} \\
\midrule
IJP       & 50  & 0.91 & 0.96 \\
GCG       & 50  & 0.98 & 0.88 \\
SAA       & 50  & 1.00 & \textbf{1.00} \\
AutoDAN   & 50  & 0.97 & 0.72 \\
PAIR      & 50  & 0.77 & 0.58 \\
DrAttack  & 50  & 0.97 & \textbf{1.00} \\
Puzzler   & 50  & 0.99 & \textbf{1.00} \\
Zulu      & 50  & 0.99 & \textbf{1.00} \\
Base64    & 50  & 0.97 & \textbf{1.00} \\
\midrule
\textbf{JB-GCG}    & \textbf{164} & \textbf{$\sim$0\%} & \textbf{100\%} \\
\bottomrule
\end{tabular}
\end{table}


\subsection{Adaptive Attack on RTV}
\label{sec:rtv_adaptive}

We evaluate RTV under adaptive adversaries that have full white-box knowledge of the defense (the fingerprint definition, calibration statistics ($\boldsymbol{\mu}_+$, $\boldsymbol{\Sigma}_+^{-1}$), the detection threshold $T$). This is the strongest possible passive threat model for a representation-level defense.

\medskip
\noindent
\textbf{Single-Layer Adaptive Attack.}
We modify the JB-GCG objective to include a fingerprint evasion term that directly targets the Mahalanobis distance:
\begin{equation}
    \mathcal{L}_{\text{RTV-GCG}}(x) = -\cos(r_{18}, h_{18}) + \frac{\lambda_M}{d_+^2(\mathbf{f}(x)) + \epsilon},
    \label{eq:rtv_gcg}
\end{equation}
where $d_+^2(\mathbf{f}(x))$ is the squared Mahalanobis distance from the fingerprint to the harmless cluster and $\epsilon = 10^{-6}$. The first term suppresses refusal at layer~18 (as in JB-GCG), while the second term pushes the fingerprint toward the harmless distribution to evade RTV detection. The objective is fully differentiable. We sweep $\lambda_M \in \{0.1, 0.5, 1.0, 2.0, 5.0\}$ on a 20-prompt pilot set and select $\lambda_M = 0.1$ for full evaluation, since this case yielded the highest ASR on the pilot set, with a budget of 600 iterations per prompt.

\medskip
\noindent
\textbf{Multi-Layer Adaptive Attack.}
Because RTV monitors layers 18, 25, and 32, we also test a stronger variant that suppresses refusal at multiple layers simultaneously:
\begin{equation}
    \mathcal{L}_{\text{ML-GCG}}(x) = -\sum_{l \in \{18, 32\}} \alpha_l \cdot \cos(r_l, {h}_l) + \frac{\lambda_M}{d_+^2(\mathbf{f}(x)) + \epsilon},
    \label{eq:ml_gcg}
\end{equation}
with $\alpha_{18} = \alpha_{32} = 1.0$ and $\lambda_M = 0.1$. This variant targets layers~18 and~32 specifically, the two layers that the pilot-sweep on layer configurations identifies as most discriminative. The rationale is that suppressing refusal at the deeper layer may prevent the inter-layer amplification that creates the fingerprint tension JB-GCG exploits.

\medskip
\noindent
\textbf{Results.}
Table~\ref{tab:adaptive_results} reports the results under SORRY-Bench evaluation. The strongest adaptive attack (ML-GCG) achieves 7.0\% ASR against RTV, compared with 53.4\% for JB-GCG against JBShield-D, a 7.6$\times$ reduction. This comes at 13$\times$ computational cost to the attacker. The average prompt exhausts most of the 600-iteration budget (514.5 iterations) due to the conflicting optimization objectives.

\begin{table}
\caption{Adaptive attack results against RTV, evaluated by SORRY-Bench. JB-GCG baseline (first row) is evaluated against JBShield-D for comparison. Cost is measured in iteration count. ML-GCG achieves 7.0\% ASR against RTV, (53.4\% for JB-GCG against JBShield-D) at 13$\times$ computational cost.}
\label{tab:adaptive_results}
\centering
\resizebox{\columnwidth}{!}{%
\begin{tabular}{lccccc}
\toprule
\textbf{Attack} & \textbf{Defense} & \textbf{ASR} & \textbf{Avg. Iter.} & \textbf{Cost} \\
\midrule
JB-GCG              & JBShield-D & 53.4\% & 39.6   & 1$\times$ \\
JB-GCG              & RTV        & 0.0\%  & 39.6    & 1$\times$ \\
RTV-GCG (Eq.~\ref{eq:rtv_gcg})  & RTV & 4.0\%  & 526.4   & 13.3$\times$ \\
ML-GCG (Eq.~\ref{eq:ml_gcg})    & RTV & \textbf{7.0}\%  & \textbf{514.5} & \textbf{13.0}$\times$ \\
\bottomrule
\end{tabular}}
\end{table}



\medskip
\noindent
\textbf{The evasion bottleneck.}
Across both adaptive attacks (200 total prompts), zero prompts achieved evasion without also producing a jailbreak response. Every prompt that evaded RTV also suppressed refusal. The converse does not hold. This asymmetry demonstrates that evasion is the binding constraint. The attacker can suppress refusal cheaply (95\% jailbreak rate) but cannot simultaneously control the 15-dimensional fingerprint to match the harmless distribution. The fundamental conflict is between the two terms in Equations~\ref{eq:rtv_gcg} and~\ref{eq:ml_gcg}. Matching the harmless fingerprint requires positive refusal-direction alignment at layer~18 ($F_{18,:} > 0$), which directly opposes the refusal-suppression term that pushes $\mathbf{h}_{18}$ away from $\mathbf{r}_{18}$.



\section{Discussion}
\label{sec:discussion}

\medskip
\noindent
\textbf{Why RTV is structurally harder to attack than JBShield.}
JBShield's AND-gate gives the attacker a binary escape. The attacker can suppress either the toxic or jailbreak concept, and the entire defense falls. JB-GCG exploits this by driving the toxic concept below threshold while leaving the jailbreak concept active. RTV eliminates this structural vulnerability. Its Mahalanobis distance operates over a 15-dimensional fingerprint with no conjunctive decomposition. The attacker must simultaneously match the legitimate distribution across all 15 dimensions, each corresponding to a specific layer-position combination's alignment with the refusal direction. Our adaptive attack evaluation confirms this. Across 200 prompts (Section~\ref{sec:rtv_adaptive}), the strongest attacker achieves only 7\% ASR at 13$\times$ computational cost, with zero evasion-only outcomes. The bottleneck is entirely on evasion, not jailbreaking.

\medskip
\noindent
\textbf{The fingerprint tension as a defense mechanism.}
The core reason RTV resists adaptive attack is a fundamental conflict between the attacker's two objectives. Suppressing refusal requires pushing the layer-18 hidden state away from the refusal direction, producing negative or near-zero cosine similarity at that layer. Matching the harmless fingerprint requires the opposite, positive cosine similarity across all layers and positions. The attacker cannot satisfy both simultaneously for most prompts. The 7\% of prompts where the attack succeeds represent cases where this tension is naturally weak. These are the queries whose representations are inherently closer to the harmless distribution. This suggests that the tension is a property of the prompt content, not a parameter the attacker can optimize away. The average iterations for successful prompts (181 vs.\ 514.5 overall for ML-GCG; 70.5 vs.\ 526.4 for RTV-GCG) confirm that successful evasions converge relatively quickly on ``easy'' prompts, while the majority of prompts exhaust the iteration budget without finding a feasible solution.

\medskip
\noindent
\textbf{Refusal directions are layer-local.}
The cosine similarity analysis in Section~\ref{sec:rtv_motivation} shows that JB-GCG prompts occupy an ambiguous distinctive zone at layer~18 but amplify toward harmful alignment at layer~32 (Figure~\ref{fig:cos-layers3_combined}). This layer-dependent behavior implies that optimizing the fingerprint at one layer provides limited control over the fingerprint at another layer, because the intermediate transformer blocks impose nonlinear transformations that do not preserve directional structure. 

\medskip
\noindent
\textbf{Connection to prior work on representation-level defenses.}
Bailey et al.~\cite{bailey2026obfuscated} demonstrate that Mahalanobis-based detectors operating on raw activations can be bypassed by obfuscation attacks that steer activations into the harmless region while maintaining harmful generation. RTV differs in a key respect. It operates on \textit{refusal-direction cosine similarities}, not raw activations. This is a lower-dimensional, more constrained feature space where the attacker is limited by the tension described above. Whether this distinction provides robustness against obfuscation-style attacks beyond GCG-style suffix optimization is an open question. We note, however, that Bailey et al.'s attack~\cite{bailey2026obfuscated} requires embedding-space access, which is a different threat model than the token-level discrete optimization we evaluate against. Extending RTV's adaptive evaluation to embedding-space attackers is an important direction for future work.

\subsection{Limitations}

\medskip
\noindent
\textbf{Single model.}
All experiments are conducted on Llama-3-8B. While the refusal-direction phenomenon has been observed across 13+ open-source models~\cite{arditi2024refusal}, we have not verified that RTV's fingerprint separation and detection performance generalize to other architectures or model scales. The layer selection (18, 25, 32) is specific to Llama-3-8B's 32-layer architecture and would need re-calibration for models with different depths.

\medskip
\noindent
\textbf{PAIR and naturalistic attacks.}
RTV achieves only 58\% detection on PAIR, which generates linguistically fluent jailbreak prompts via iterative LLM refinement. PAIR prompts produce representations that are inherently close to the harmless distribution in fingerprint space, making them difficult for any representation-level detector. JBShield also struggles with PAIR (77\% accuracy), suggesting this is a shared limitation of the defense paradigm rather than an RTV-specific weakness. Defenses that combine representation-level signals with semantic analysis of prompt content may be necessary to address this class of attacks.

\medskip
\noindent
\textbf{Nonzero adaptive ASR}
The strongest adaptive attack achieves 7\% ASR. While this is a substantial reduction from JB-GCG's 53.4\% against JBShield, it is not zero. The successful evasions correspond to prompts where the tension between refusal suppression and fingerprint evasion is naturally weak. Closing this gap may require adaptive thresholding, prompt-dependent fingerprint normalization, or integration with complementary defense mechanisms.

\medskip
\noindent
\textbf{Static calibration.}
RTV's calibration statistics ($\boldsymbol{\mu}_+$, $\boldsymbol{\Sigma}_+$, threshold $T$) are fixed after calibration. While our stability analysis shows low sensitivity to calibration randomness (AUROC $0.994 \pm 0.003$), the defense does not adapt to distributional shift in the input population. In deployment scenarios where the distribution of user prompts evolves over time, periodic recalibration may be necessary.


\section{Conclusion}
\label{sec:conclusion}

We study the robustness of representation-level jailbreak defenses under adaptive threat models using JBShield as a target. Our adaptive attack, JB-GCG, combines refusal-direction suppression with detector-aware optimization to exploit JBShield's detection rule, achieving up to 53.4\% ASR where JBShield reports 0\% against standard GCG. We observe that the attack consistently breaks the AND-gate by suppressing the toxic concept while leaving the jailbreak concept active. This demonstrates that strong performance against non-adaptive baselines does not imply robustness against adaptive adversaries.

Analyzing why JB-GCG succeeds reveals a multi-layer representation signature. Adversarial prompts occupy a distinctive zone in the refusal-direction based fingerprint space. This observation motivates Representation Trajectory Verification (RTV), a defense that monitors refusal-direction cosine similarity across multiple layers and token positions and detects outliers via Mahalanobis distance in the 15-dimensional fingerprint space. RTV achieves 0.99 AUROC against JB-GCG without requiring any jailbreak examples for calibration, and generalizes to five of nine standard attack types with perfect detection.

Under the strongest adaptive attack, with full white-box knowledge of the defense, including calibration statistics and detection threshold, the attacker achieves only 7\% ASR at 13$\times$ computational cost. Our results suggest that multi-layer representation consistency provides a more robust foundation for jailbreak detection than single-layer concept similarity. We note that challenges remain for naturalistic attacks and stronger threat models involving embedding-space optimization.

\bibliographystyle{ACM-Reference-Format}
\bibliography{software}

\appendix

\section{Model Ablation} \label{app:chat_ablation}

\begin{figure}
    \centering
    \includegraphics[width=\linewidth]{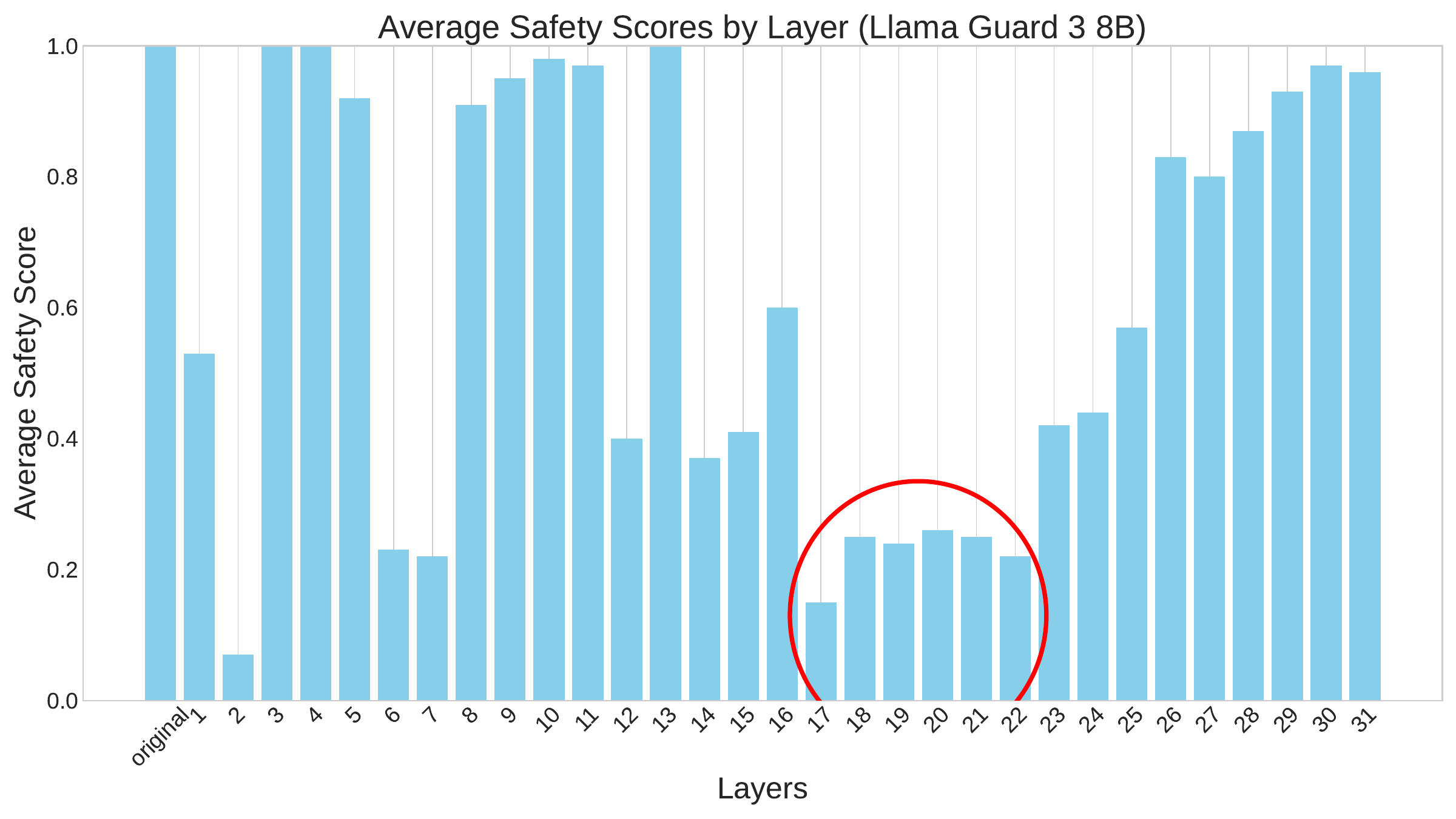}
    \caption{Layer-wise refusal-direction ablation on Llama-3 8B, evaluated using Llama-Guard-3 8B. Lower safety scores indicate stronger harmfulness after ablation.}
    \label{fig:ablated_chat_llama3}
\end{figure}

For each layer, we extract a refusal direction and then ablate layers by subtracting the projection of the refusal direction from the residual stream. We generate responses from the ablated model and evaluate them with Llama-Guard-3 8B. In this evaluation, a harmless response receives a score close to 1, while a harmful response receives a score close to 0. Figure~\ref{fig:ablated_chat_llama3} shows that layers 17--22 yield the lowest safety scores after ablation, indicating that refusal-related behavior is especially sensitive in this range.

\section{Joint-GCG Cosine-similarity Trajectories}
\label{app:joint_gcg_intermediate}

\begin{figure}
    \centering
    \begin{subfigure}{\linewidth}
        \centering
        \includegraphics[width=\linewidth]{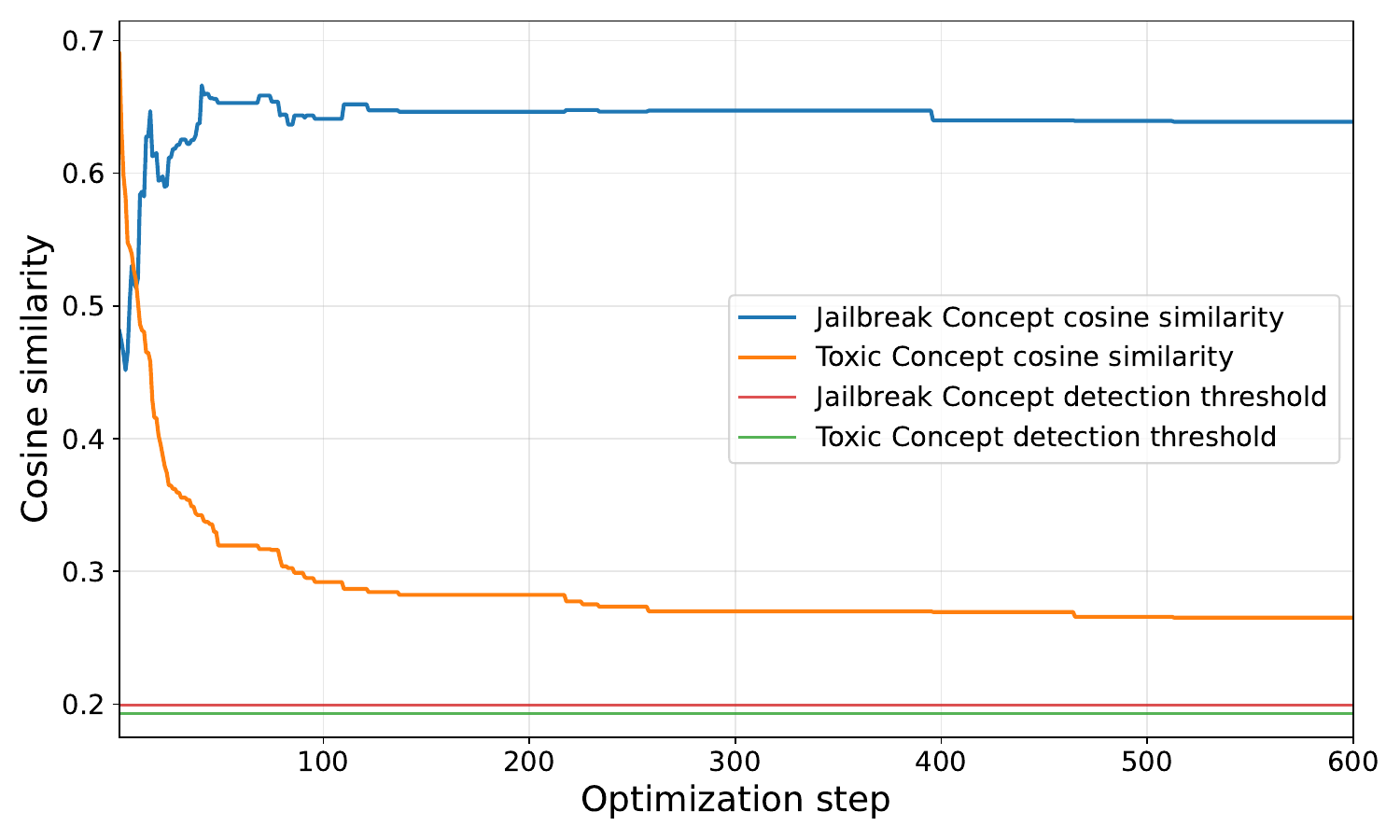}
        \caption{$\alpha=1.0,\ \beta=0.0$}
        \label{fig:joint_toxic_only}
    \end{subfigure}
    \hfill
    \begin{subfigure}{\linewidth}
        \centering
        \includegraphics[width=\linewidth]{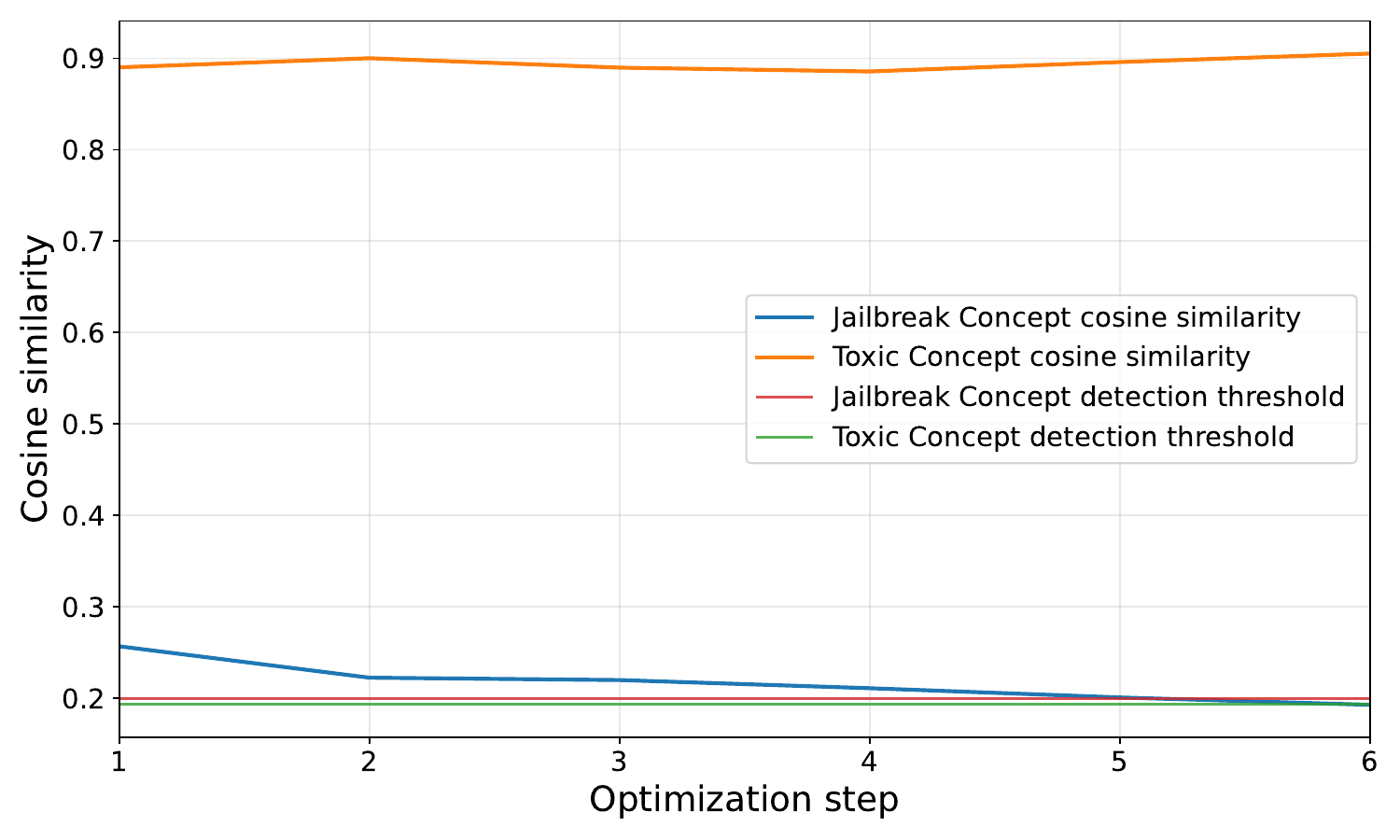}
        \caption{$\alpha=0.0,\ \beta=1.0$}
        \label{fig:joint_jb_only}
    \end{subfigure}
    \caption{Joint-GCG cosine-similarity trajectories under extreme weighting. (a)~Targeting toxic concept only: the toxic score decreases but remains above threshold, while the jailbreak score rises sharply. (b)~Targeting jailbreak concept only: evasion is achieved within a few steps, but the toxic score remains high and no harmful outputs are produced.}
    \label{fig:alternating_gcg_combined}
\end{figure}


Figure~\ref{fig:joint_toxic_only} illustrates the cosine similarity trajectories for $\alpha=1.0$ and $\beta=0.0$. Under this configuration, the toxic concept score decreases throughout optimization. The jailbreak concept score remains high and does not fall below its threshold. This configuration fails to produce a suffix since both concept similarities stay above their thresholds. Figure~\ref{fig:joint_jb_only} shows the corresponding trajectories for $\alpha=0.0$ and $\beta=1.0$, where jailbreak concept cosine similarity descends below its respective detection threshold during optimization in a few optimization steps.

\section{Cluster Separation}
\label{app:clusters}

\begin{figure}
    \centering
    \begin{subfigure}{\linewidth}
        \centering
        \includegraphics[width=\linewidth]{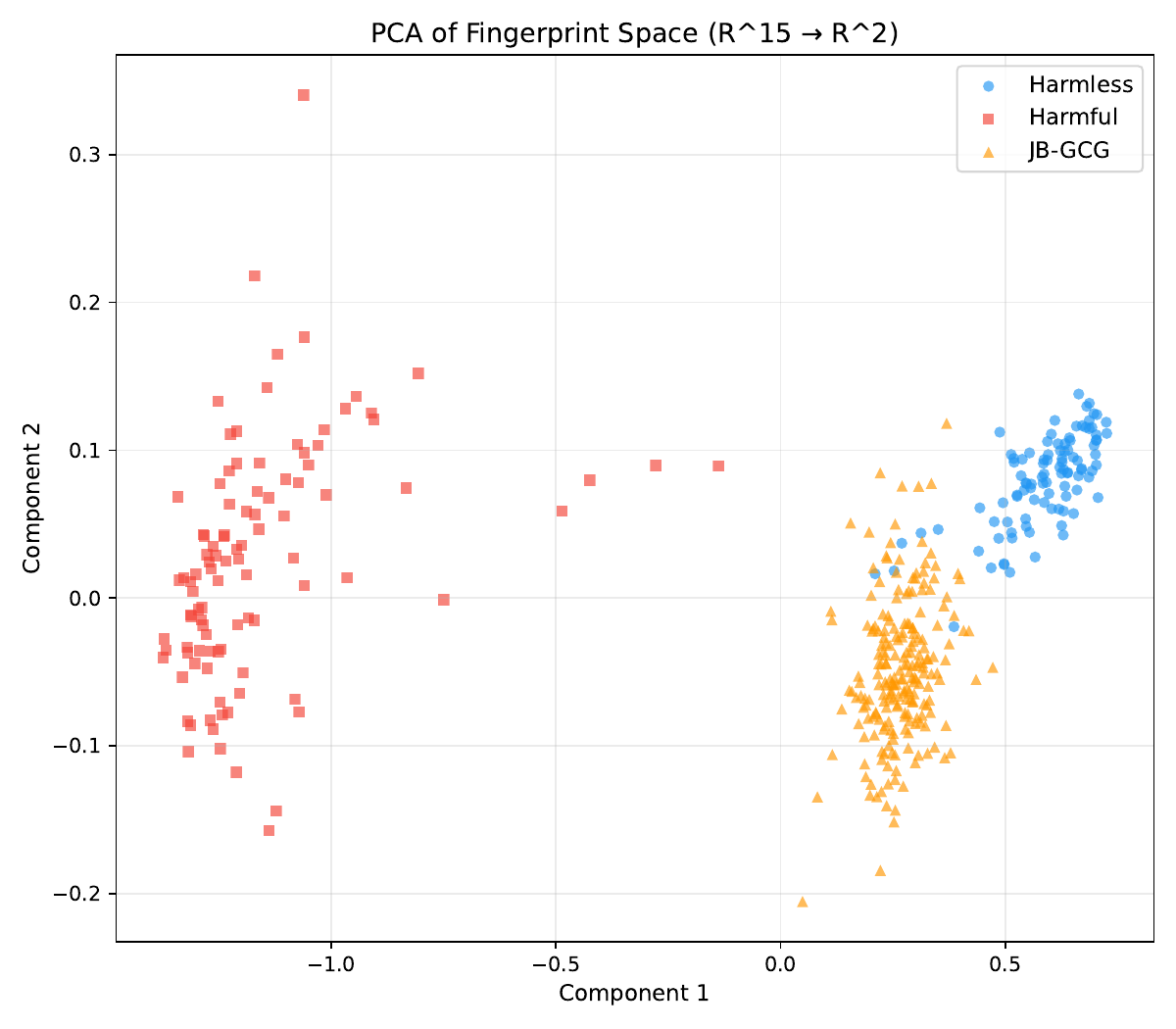}
        \caption{PCA projection.}
        \label{fig:sca-pca}
    \end{subfigure}
    \hfill
    \begin{subfigure}{\linewidth}
        \centering
        \includegraphics[width=\linewidth]{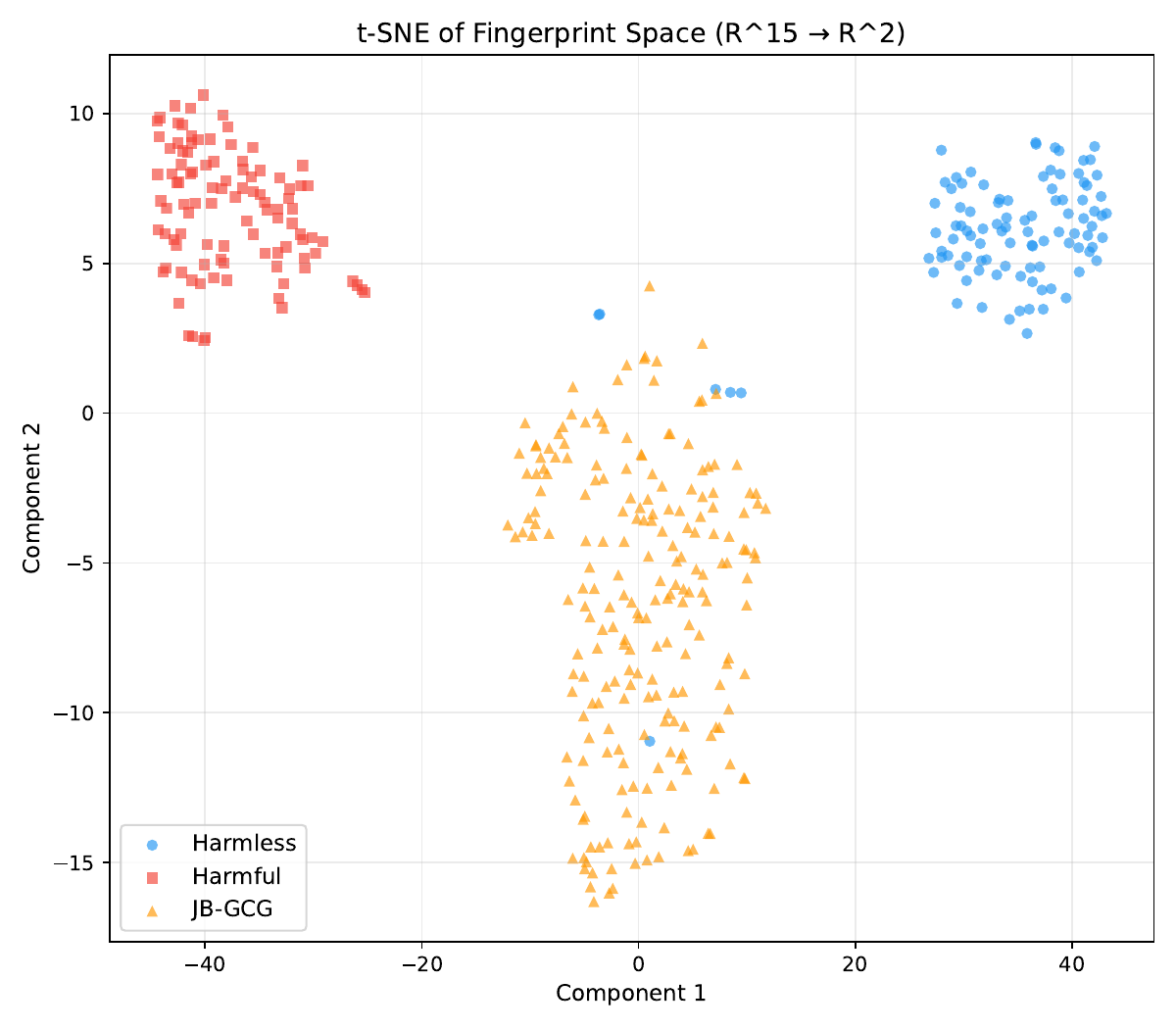}
        \caption{t-SNE projection.}
        \label{fig:sca-sne}
    \end{subfigure}
    \caption{PCA and t-SNE projections of the 15-dimensional fingerprint space. Three categories form distinct clusters, with JB-GCG occupying the intermediate region between harmless and harmful.}
    \label{fig:sca-combined}
\end{figure}

Figure~\ref{fig:sca-combined} confirms that the distinctive zone pattern translates to per-prompt cluster separation. In PCA projection of the 15-dimensional fingerprint space, the three categories form distinct clusters, harmful on the far left, harmless on the far right, and JB-GCG in between. While JB-GCG and harmless are adjacent, they remain separable with a visible gap between the bulk of each cluster. The t-SNE projection shows even sharper separation, with no JB-GCG point falling within either legitimate cluster. This separation motivates a statistical outlier detector in the fingerprint space. The PCA projection shows the separation exists in the original 15-dimensional space, not just in a nonlinear embedding.

\section{Calibration Stability}
\label{app:calibration_stability2}
We verify that RTV's performance is not sensitive to the specific calibration data. Table~\ref{tab:rtv_stability} reports results across five random calibration splits (each using 100 harmless + 100 harmful prompts drawn from Alpaca~\cite{alpaca} and HarmBench~\cite{mazeika2024harmbench}, respectively), evaluated on the fixed test set. AUROC varies by only 0.003 across splits, while TPR remains at or above 99.4\%.

\begin{table}
\caption{RTV stability across random calibration splits.}
\label{tab:rtv_stability}
\centering
\begin{tabular}{lcc}
\toprule
 & \textbf{AUROC} & \textbf{TPR @ 5\% FPR} \\
\midrule
Mean $\pm$ Std & $0.994 \pm 0.003$ & $99.9\% \pm 0.2\%$ \\
\bottomrule
\end{tabular}
\end{table}

\section{Calibration-Size Sensitivity}
\label{app:calibration_size}

We evaluate RTV across calibration sizes $N \in \{10, 20, 30, 50, 100\}$ per class. The threshold $T$ is recalibrated at each $N$ to target 5\% FPR on the combined harmless and harmful test set. As shown in Table~\ref{tab:calibration_size}, RTV achieves $\geq$97\% detection for all $N \geq 20$. At $N=10$, the Ledoit-Wolf covariance estimate is unreliable in 15 dimensions, producing an inflated threshold (11.03) and reduced detection (79.3\%). The threshold stabilizes around 4.5--5.2 for $N \geq 20$, and FPR tracks close to the 5\% target across both harmless and harmful test sets.

\begin{table}
\caption{Calibration-size sensitivity of RTV. Detection rate is on JB-GCG ($n=164$). FPR-h and FPR-hm denote false positive rates on harmless and harmful test prompts, respectively.}
\label{tab:calibration_size}
\centering
\small
\begin{tabular}{ccccc}
\toprule
\textbf{$N$/class} & \textbf{Threshold $T$} & \textbf{Det.\ Rate} & \textbf{FPR-h} & \textbf{FPR-hm} \\
\midrule
10  & 11.03 & 79.3\% (130/164) & 2.0\% & 8.0\% \\
20  & 5.19  & 100.0\% (164/164) & 5.0\% & 5.0\% \\
30  & 4.80  & 97.0\% (159/164) & 4.0\% & 6.0\% \\
50  & 4.49  & 100.0\% (164/164) & 7.0\% & 3.0\% \\
100 & 4.71  & 100.0\% (164/164) & 7.0\% & 3.0\% \\
\bottomrule
\end{tabular}
\end{table}

\section{Ablation Studies}
\label{app:ablations}

To understand how the performance varies with fingerprint dimensions beyond the chosen configuration (3 layers, 5 token positions), we perform further experiments with results summarizes in Tables~\ref{tab:layer_ablation} and \ref{tab:position_ablation}.

\subsection{Layer Ablation}

\begin{table}
\caption{RTV detection with different layer combinations. Single layer configurations use 5-dimensional fingerprint space. Two-layer configuration uses 10-dimensional fingerprint space. Three and five layer configuration use 15 and 25 dimensional fingerprint space, respectively.}
\label{tab:layer_ablation}
\centering
\begin{tabular}{lccc}
\toprule
\textbf{Layers} & \textbf{Dim} & \textbf{JB-GCG Det.} & \textbf{AUROC} \\
\midrule
\{18\}              & 5  & 97.56\% & 0.9924 \\
\{25\}              & 5  & 68.90\% & 0.9463 \\
\{32\}              & 5  & 95.12\% & 0.9874 \\
\{18, 32\}          & 10 & 99.39\% & 0.9964 \\
\{18, 25, 32\}      & 15 & 100.0\% & 0.9946 \\
\{18, 22, 25, 29, 32\} & 25 & 100.0\% & 0.9962 \\
\bottomrule
\end{tabular}
\end{table}

Table~\ref{tab:layer_ablation} shows RTV detection performance across different layer combinations. Layer~18 alone achieves 0.9924 AUROC, confirming it is the primary detection layer, the layer where JB-GCG directly optimizes refusal suppression. Adding layer~32 improves detection from 97.56\% to 99.39\% (AUROC 0.9964). Layer~25 contributes minimally: adding it to \{18, 32\} improves detection to 100\% but slightly decreases AUROC (0.9946 vs.\ 0.9964), likely due to increased covariance estimation noise. Five layers provide no benefit over three.

\subsection{Position Ablation}

\begin{table}
\caption{RTV detection with different token position counts ($K=3$ layers).}
\label{tab:position_ablation}
\centering
\begin{tabular}{lccc}
\toprule
\textbf{Positions} & \textbf{Dim} & \textbf{JB-GCG Det.} & \textbf{AUROC} \\
\midrule
\{0\}                   & 3  & 92.68\% & 0.9761 \\
\{$-$1, 0\}             & 6  & 96.34\% & 0.9853 \\
\{$-$2, $-$1, 0\}       & 9  & 99.39\% & 0.9917 \\
\{$-$4, ..., 0\}        & 15 & 100.0\% & 0.9946 \\
\bottomrule
\end{tabular}
\end{table}

As shown in Table~\ref{tab:position_ablation}, each additional token position adds meaningful discriminative power, with improvement from 0.9761 to 0.9946 AUROC. The gain is consistent across steps, validating the causal-attention argument. The adversarial suffix's influence decays with distance from the last token, so earlier positions carry independent information about the uninfluenced prompt content.







\end{document}